\documentclass[aps,prl,superscriptaddress,11pt]{revtex4-2}
\usepackage[colorlinks,allcolors=blue]{hyperref}
\usepackage[normalem]{ulem}
\usepackage{amsmath,amssymb}
\usepackage{mathrsfs,CJK,newtxtext,newtxmath}
\usepackage{xcolor,graphicx,natbib}
\begin{document}

\title{Quantum dynamics of topological strings in a frustrated Ising antiferromagnet}
\author{Zheng Zhou}
\affiliation{Department of Physics and State Key Laboratory of Surface Physics, Fudan University, Shanghai 200438, China}
\affiliation{Perimeter Institute for Theoretical Physics, Waterloo, Ontario N2L 2Y5, Canada}
\author{Changle Liu}
\affiliation{Department of Physics, and Center of Quantum Materials and Devices, Chongqing University, Chongqing, 401331, China}
\affiliation{Shenzhen Institute for Quantum Science and Technology and Department of Physics, Southern University of Science and Technology, Shenzhen 518055, China}
\author{Zheng Yan }
\email{zhengyan@hku.hk}
\affiliation{Beihang Hangzhou Innovation Institute, Yuhang, Hangzhou 310023, China}
\affiliation{Department of Physics and HKU-UCAS Joint Institute of Theoretical and Computational Physics, The University of Hong Kong, Pokfulam Road, Hong Kong}
\author{Yan Chen }
\email{yanchen99@fudan.edu.cn}
\affiliation{Department of Physics and State Key Laboratory of Surface Physics, Fudan University, Shanghai 200438, China}
\affiliation{Collaborative Innovation Center of Advanced Microstructures, Nanjing 210093, China}
\author{Xue-Feng Zhang }
\email{zhangxf@cqu.edu.cn}
\affiliation{Department of Physics, and Center of Quantum Materials and Devices, Chongqing University, Chongqing, 401331, China}
\affiliation{Chongqing Key Laboratory for Strongly Coupled Physics, Chongqing University, Chongqing 401331, China}

\begin{abstract}
We investigate the quantum dynamics of the antiferromagnetic transverse field Ising model on the triangular lattice through large-scale quantum Monte Carlo simulations and stochastic analytic continuation. This model effectively describes a series of triangular rare-earth compounds, for example, $\mathrm{TmMgGaO}_4$. At weak transverse field, we capture the excitations related to topological quantum strings, which exhibit continuum features described by XY chain along the strings and those in accord with `Luttinger string liquid' in the perpendicular direction. The continuum features can be well understood from the perspective of topological strings. Furthermore, we identify the contribution of strings from the excitation spectrum. Our study provides characteristic features for the experimental search for string-related excitations and proposes a theoretical method to pinpoint topological excitations in the experimental spectra.

\textbf{Keywords}: quantum strings, rare-earth compounds, frustrated magnetism, transverse field Ising model
\end{abstract}

\maketitle
\clearpage

\section{Introduction}

Frustrated magnets provide an ideal playground for exotic emergent phenomena~\cite{frus_1,frus_2}.
In many cases, gauge structures can emerge due to local constraints imposed by competing interactions, giving rise to exotic states of matter like quantum spin liquid with fractionalised excitations and long-range entanglement~\cite{gauge_1,gauge_2,gauge_3,gauge_4,topological2021yan,yan2022qdm,yan2020improved}.
Excitations violating such local constraints are frequently referred to as spinons, which are the gauge charges in the emergent gauge theory~\cite{spinice_1,gang_chen,RK2021}.
Spinons are point-like topological defects since a single spinon cannot be created/annihilated locally and such processes must be involved in pairs.
Meanwhile, gauge theories also support string-like excitations, which connect a pair of spinons with opposite charges and mediate the effective interaction between them.
While the spinon pair is annihilated along a non-contractible closed trajectory, such string becomes topological in nature, which we dub as `topological string'~\cite{frank,yan2022targeting}.

In contrast to spinon excitations that are typically located at higher energies, quantum strings are basic degrees of freedom participating in low-energy physics.
%
Their dynamics reflects the fluctuations of the background gauge field~\cite{tfim_str_1,tfim_str_2}.
Quantum strings are 1d objects in space that can be described as two-dimensional worldsheets in spacetime~\cite{ising_3d}.
As emergent topological defects with finite dimensions, strings have much more internal degrees of freedom (corresponding to their vibration) and are thus expected to bring much richer physics.
Strings appear in broad contexts in condensed matter physics, such as in high-temperature superconductors~\cite{zaanen_89,zaanen_96,Machida,zaanen_00,zaanen_01}, quantum spin ice~\cite{spinice_2,spin_ice,dirac_string} and incommensurability~\cite{SC_6,zaanen_89,incom_1}.
They can even be directly observed via the ultra-cold atoms in the optical lattice~\cite{str_cold_1,str_cold_2}.
However, in contrast to well-studied point-like excitations, strings are paid much less attention.

The quest for exploring string dynamics is also motivated by the rapid progress on frustrated rare-earth materials.
Recently, a number of rare-earth pyroclore, Kagom\'e and triangular lattice magnets have been evidenced to be described by frustrated quantum Ising models~\cite{dun2018quantum,tmgo_neu,tmgo_uud,tmgo_kt,tmgo_model,mat_lgt_1} which exhibit emergent gauge structures.
For the triangular lattice, a large class of materials are described by the transverse field Ising model~(TFIM)~\cite{tmgo_model,tmgo_neu}, a typical model that can be understood from the perspective of topological strings~\cite{tfim_str_1,tfim_str_2,str_cold_2}:
String picture can be utilised to understand its phase diagram~\cite{tfim_str_1,tfim_str_2} as well as to depict the quantum fluctuations therein~\cite{flux,orland_1993,orland_1991,orland_1992}.
Among these materials, a typical one is $\mathrm{TmMgGaO}_4$~(TMGO).
The spin excitation spectra was recently discovered in the neutron scattering experiment~\cite{tmgo_neu}, and the so-called `roton' excitation was claimed to relate to the BKT physics~\cite{tmgo_kt,tfim_qmc_3,referee_2}.
However, direct signatures of string excitations are absent from the neutron spectrum.

In this manuscript, we study a generic antiferromagnetic TFIM on the triangular lattice relevant to triangular lattice rare-earth compounds.  Through large-scale quantum Monte Carlo (QMC) simulations~\cite{qmc_1}, we focus on the dynamical signatures of quantum string excitations.
At weak transverse field, we find continuum excitations at low energies that accords with the internal vibration and mutual interactions of quantum strings.
In particular, the internal vibration matches well with the continuous spectrum of spin-$1/2$ XY chain, while the string interactions induce excitations exhibiting features of Luttinger liquid~\cite{luttinger}.
It should be noted that these continuum spectroscopic properties go beyond the effective field theory, where for the latter only coherent spin waves were predicted at low energies.
Our work provides insight into understanding the dynamical properties of triangular quantum Ising antiferromagnet such as TMGO.

\section{Results}

\subsection{Starting from rare-earth compounds}

We begin with the nearest-neighbour spin-$1/2$ TFIM relevant to triangular rare-earth compounds
\begin{equation}
    H=J\sum_{\langle ij\rangle}S_i^zS_j^z-\Delta\sum_iS_i^x.
    \label{eq_hmt}
\end{equation}
Here $S$ is the effective spin-$1/2$ moment acting on the lowest two crystal field levels of the non-Kramers ion.
$J$ denotes the nearest neighbour antiferromagnet coupling and the $\Delta$ is the energy splitting between the lowest crystal-field levels.
%
The origin of the crystal field splitting term $\Delta$ depends on the nature of crystal field levels. For non-Kramers doublet systems, the crystal field degeneracy is protected by the $C_3$ on-site point group symmetry, therefore $\Delta$ is induced by breaking such symmetry, \emph{e.g.}, by applying in-plane uniaxial pressure. For TMGO and some other Tm-based materials, the lowest two crystal field levels form a nondegenerate dipole-multipole doublet. The lowest crystal fields exhibit intrinsic splitting  $\Delta$ since they carry the singlet representation of the point group.
To quantitatively compare with the TMGO experiment, we also introduce a next-nearest-neighbour interaction $H_{J'}=J'\sum_{\langle\langle ij\rangle\rangle}S_i^zS_j^z$ perturbation term to the Hamiltonian~\cite{tmgo_model}.
The effective model is illustrated in Fig.~\ref{fig_2}c inset.

\begin{figure}[t!]
    \centering
    \includegraphics[width=.7\linewidth]{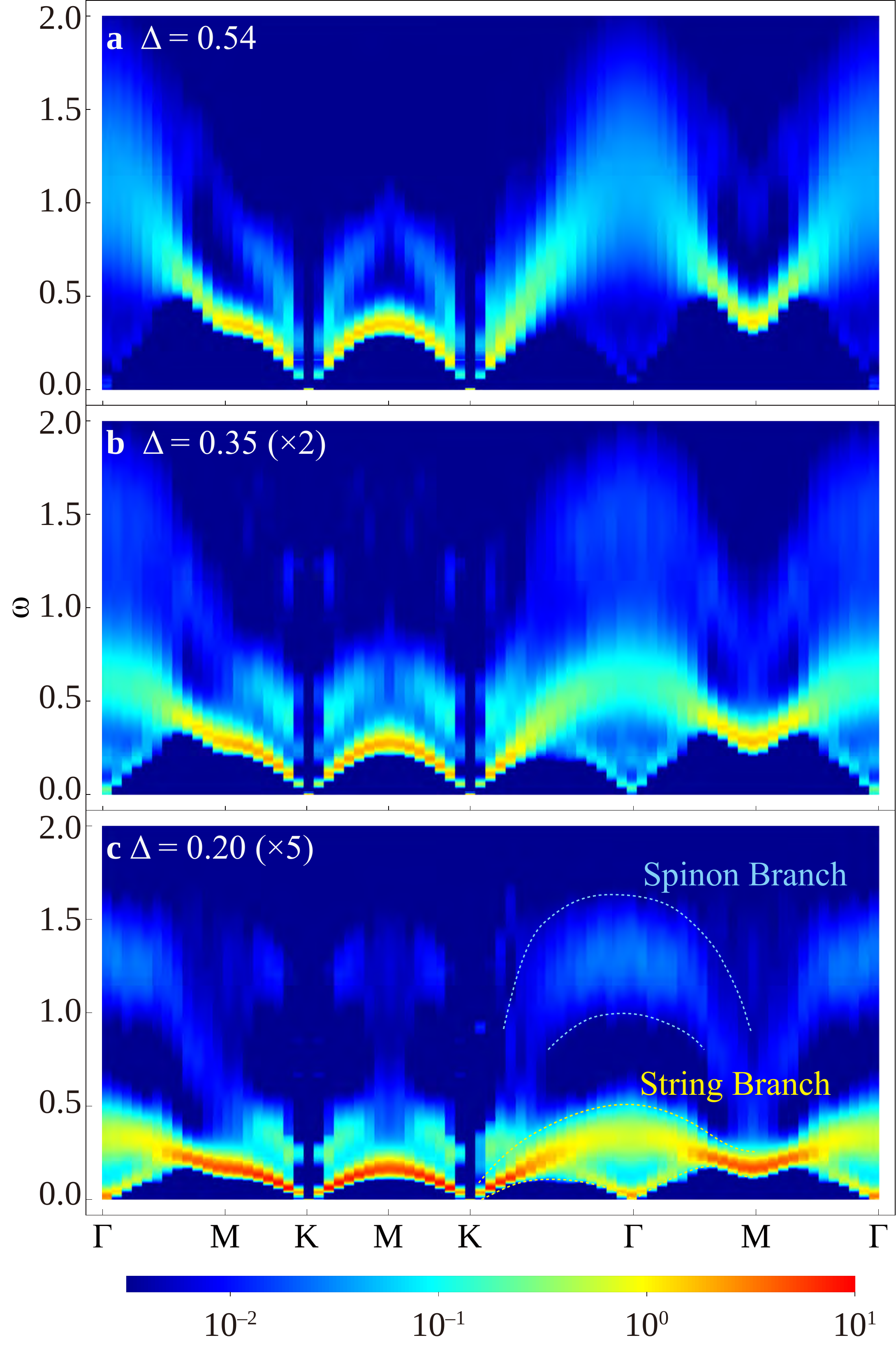}
    \caption{\textbf{The spectra at different transverse field $\Delta$.} \textbf{a}, The parameter $\Delta=0.54$ and $J'=0.05$ corresponds to the material TMGO. \textbf{b}, $\Delta=0.35$, $J'=0.035$ and \textbf{c}, $\Delta=0.20$, $J'=0.020$ are for lower energy splitting. The circled areas in \textbf{c} demonstrate the splitting of two branches. The path taken in the Brillouin zone is specified in the inset of \textbf{a}. The spectra are plotted in logarithmic scale to accommodate more details. For the plots in linear scale, see Supplementary Figure 2.}
    \label{fig_1}
\end{figure}

The best-fit parameters for TMGO were found in the former work~\cite{tmgo_kt} to be $J=0.99~\mathrm{meV}$, $J'=0.05~\mathrm{meV}$ and $\Delta=0.54~\mathrm{meV}$.
Hereafter we set the NN coupling as the energy unit $J=1$.
We carry out a QMC simulation~\cite{qmc_1,qmc_2,qmc_3} combined with SAC technique~\cite{sac_1,sac_2,sac_3} to measure the spin excitation spectrum.
The simulation is performed on $L\times L$ lattice ($L=24$) with periodic boundary conditions. The temperature is set  $T=1/4L=1/96$ to mimic the ground state result.
In inelastic neutron scattering experiments, what is measured is the dynamical $S^z$-$S^z$ correlator given by
\begin{equation}
\mathscr{S}^{zz}(\mathbf{q},\omega)=\frac{1}{2\pi L^2}\sum_{ij}\int_{-\infty}^{+\infty}\mbox{d}t\,e^{i\mathbf{q}\cdot(\mathbf{r}_{i}-\mathbf{r}_{j})-i\omega t}\langle S_{i}^{z}(0)S_{j}^{z}(t)\rangle.
\end{equation}
Our measured excitation spectrum is presented in Fig.~\ref{fig_1}a.

At the parameters for TMGO, the simulated result agrees well with the previous inelastic neutron scattering experiment~\cite{tmgo_neu} and the previous QMC work~\cite{tmgo_kt} (See Supplementary Figure 1).
Notice that for TMGO $\Delta$ is large where the system has been close to the clock-paramagnetic transition point~(Fig.~\ref{fig_2}f).
%
To gain a better understanding of the physics deep in the clock phase, we also measure the excitation spectra with decreased $\Delta$.
We find that with decreasing $\Delta$, the spectrum splits into two branches.
The higher-lying and lower-lying branches lie at the energy scale of $J$ and $\Delta$, respectively, see Fig.~\ref{fig_1}b,c.

\subsection{The string description}

\begin{figure}[t!]
    \centering
    \includegraphics[width=.7\linewidth]{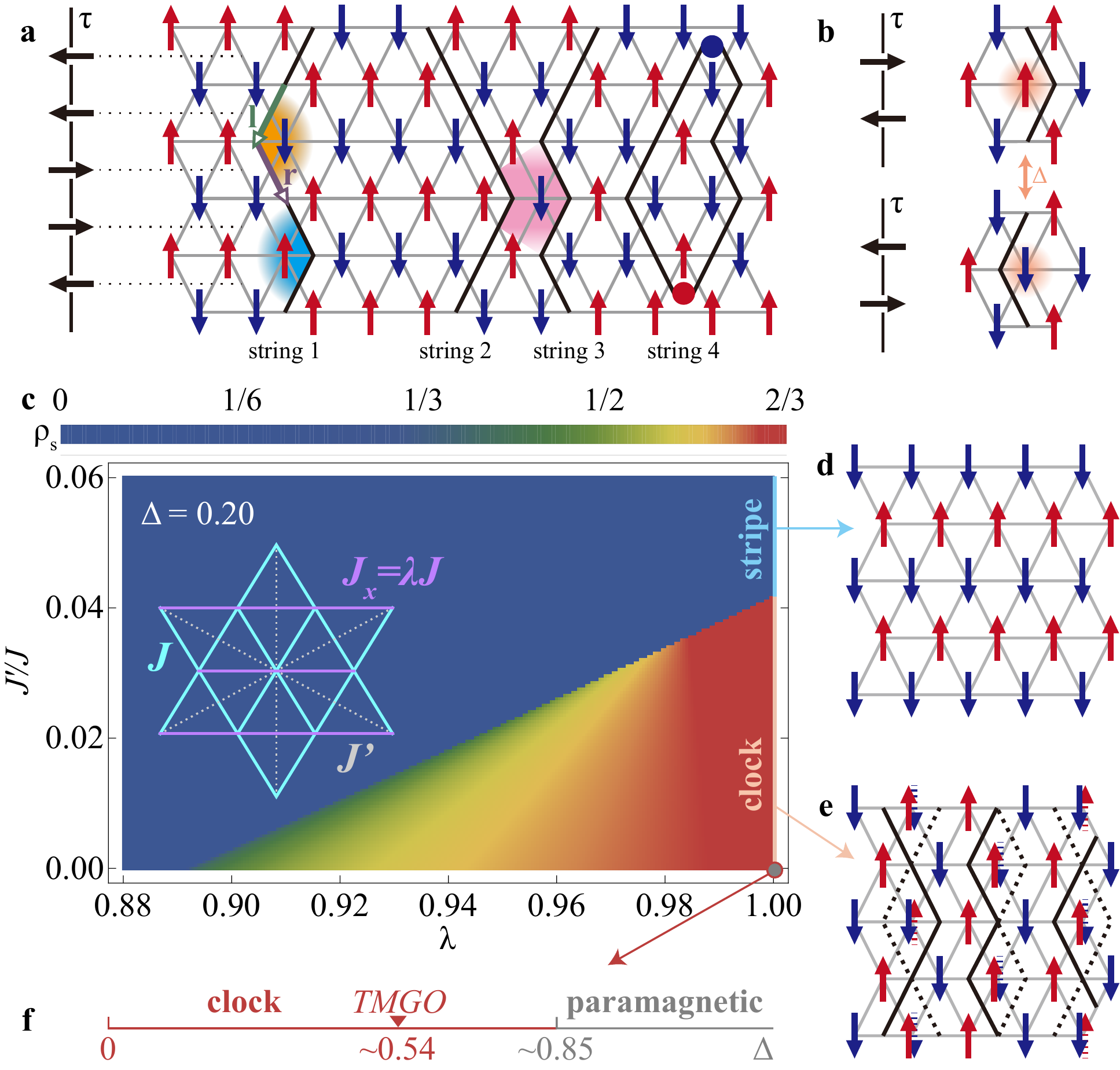}
    \caption{\textbf{An illustration of the emergent strings in frustrated TFIM.} \textbf{a}, An example of string mapping of a spin configuration. Strings~1, 2 and 3 don't involve triangle-rule-breaking configurations and are thus directional. The $\mathbf{l}$ and $\mathbf{r}$ vectors on string 1 denote the two directions of the segments. The spin chain labelled $\tau$ on the left illustrates the effective spin-$1/2$ configuration mapped from string~1. The orange and cyan shaded angles on string~1 denote a kink-antikink pair. The pink shaded area illustrates that two nearby strings induce repulsion. String 4 is a closed non-topological string connecting two spinons denoted by red and blue points. \textbf{b}, The elementary flipping process brought by the transverse field $|\mathbf{lr}\rangle\leftrightarrow|\mathbf{rl}\rangle$. The transverse field acts upon the shaded spin. \textbf{c}, The ground state string density as a function of anisotropy $\lambda$ and NNN coupling $J'$, obtained from large scale QMC simulation. Inset: an illustration of the lattice setup. \textbf{d,e}, Configuration and string correspondence of two specific phases: stripe phase~(\textbf{d}) with average string density $\langle\rho_\mathrm{s}\rangle=0$ and clock phase~(\textbf{e}) with $\langle\rho_\mathrm{s}\rangle=2/3$. The collapsed arrows denote the superposition of spin-up and down. Correspondingly, the strings are superpositions of configurations denoted by the solid and dashed lines. \textbf{f}, The phase diagram at $\lambda=1$ and $J'=0$, at different transverse field $\Delta$. }
    \label{fig_2}
\end{figure}

To better understand the two branches of excitations, we first consider the classical Ising limit $\Delta=0$. In this case, the ground state is macroscopically degenerate which must satisfy the `2-up-1-down' or `1-up-2-down' triangle-rule within each elementary triangle. Such local constraints give rise to emergent $U(1)$ gauge structures at low energies which further lead to emergent topological strings and topological sectors. Instead of understanding strings from dimers and gauge theories, here we present a more intuitive way from an alternative domain wall perspective~\cite{tfim_map_2,tfim_map_3,tfim_map_4}. We choose the stripe state (Fig.~\ref{fig_2}d) as the reference configuration. The reason we choose the stripe state as the reference state comes is that any local operation acting upon this state violates the local constraint, hence there are no low-energy excitations within the stripe bulk. With this setup, all spin configurations which obey the triangle-rule can be expressed as domains of stripe state, where domain walls form closed directed strings (Fig.~\ref{fig_2}a, strings 1--3). In other words, one selects out the bonds in $x$-direction connecting anti-parallel spins, and connects their mid-points into strings. The strings can be visualised in measurements (See Supplementary Figure 4). Under periodic boundary conditions, strings wrapping around non-contractible loops cannot be created or annihilated within finite steps of operation, which strongly reflects its topological feature in nature. Therefore the number of topological strings $N_\mathrm{s}$ can be used to label different topological sectors. On the other hand, breaking of triangle-rule corresponds to spinon excitations at the energy scale of $J$~\cite{tfim_str_1,tfim_str_2} (Fig.~\ref{fig_2}a, string 4) . This is related to the high energy excitations.

Once the transverse field $\Delta$ is tuned on, quantum dynamics are present. When $\Delta\ll J$, the low energy excitations do not mix up with the spinon excitations at a higher energy level. As shown above, the low energy quantum excitations are completely provided by the quantum strings. Here we first consider the excitations where only one string is involved, i.e., the internal vibration of a single string. The shape of a string is specified by its segments $\{\mathbf{s}_1,\dots,\mathbf{s}_{L_y}\}$, which take on the values $\mathbf{r}$ and $\mathbf{l}$ (shown in Fig.~\ref{fig_2}a). Upon the action of the transverse field, a pair of antiparallel nearest-neighbour spins are flipped, i.e.,  $|\dots\mathbf{r}\mathbf{l}\dots\rangle\leftrightarrow|\dots\mathbf{l}\mathbf{r}\dots\rangle$ (Fig.~\ref{fig_2}b). According to the first-order perturbation theory, these fluctuations can be mapped onto a spin-$1/2$
XY chain, with spin values $\tau^z=\pm 1/2$ represents $\mathbf{r}$ and $\mathbf{l}$ segments
\begin{equation}
    H_\mathrm{eff}=-\frac{\Delta}{2}\sum_{i=1}^{L_y}(\tau_i^+\tau_{i+1}^-+\tau_i^-\tau_{i+1}^+)
\end{equation}
The effective model is exactly solvable under Jordan-Wigner transformation. Therefore, one can predict a continuous spectrum due to effective Jordan-Wigner fermions excitations corresponding to kink-antikink pairs in the language of strings. This vibration also brings an energy favour $E_\mathrm{k}=-\Delta L_y/\pi$ to each string, stablizing it energetically against perturbations.

When multiple strings are considered, their interaction becomes significant to the quantum excitations. The non-crossing condition first imposes a hard-core repulsion on strings. The vibration of strings then turns this hard-core repulsion into a dynamic one~\cite{tfim_map_2}. Specifically, when two strings come close to each other (e.g., the string~2 and 3 in Fig.~\ref{fig_2}a), their vibration will be constrained, reducing the energy gain from internal vibration (See Supplementary Note 1). This loss of vibration energy can be equivalently regarded as the repulsion interaction between them. As former work~\cite{incom_1} has identified, the dynamic repulsion can be described by a power-law long-range potential. In other words, the internal vibration of strings can also generate mutual dynamic interaction among them.

\subsection{Dilute string limit}

The vibration inside the strings, together with the repulsion interaction between them, determines the low energy excitations of TFIM.
To study these two kinds of excitations separately, we turn to the dilute string limit where the strings are far apart and the interaction between them is weak, so that the internal string vibration is not affected by string interaction.
This can be parametrised by a quantity called the string density $\rho_\mathrm{s}=N_\mathrm{s}/L_x$.
The string density is determined by measuring the nearest-neighbor correlation $\langle\rho_\mathrm{s}\rangle=\frac{1}{2}\left(1-4\langle S_\mathbf{r}^zS_{\mathbf{r}+\hat{\mathbf{x}}}^z\rangle\right)$.
In particular, the stripe phase and clock phase, whose configurations and string correspondence are depicted in Fig.~\ref{fig_2}d,e, have averaged string density $\langle\rho_\mathrm{s}\rangle_\mathrm{stripe}=0$ and $\langle\rho_\mathrm{s}\rangle_\mathrm{clock}=2/3$.
The string density can be tuned through introducing various perturbations, like the nearest-neighbour exchange anisotropy $H_\lambda=(\lambda-1)J\sum_{\langle ij\rangle_x}S_i^zS_j^z$~\cite{incom_1}.
Here $\langle ij\rangle_x$ denote the NN bonds in $x$-direction~(Fig.~\ref{fig_2}c inset).
The string density $\rho_\mathrm{s}(\lambda,J')$ as a function of exchange anisotropy $\lambda$ and NNN interaction $J'$ are measured through a Monte Carlo simulation, shown in Fig.~\ref{fig_2}c (See Supplementary Note 2).
By increasing $J'$, the system enters the stripe phase from the clock phase through a first-order transition; by decreasing $\lambda$, the string density $\rho_\mathrm{s}$ decreases from $2/3$ to $0$ continuously.

\begin{figure}[t!]
    \centering
    \includegraphics[width=.7\linewidth]{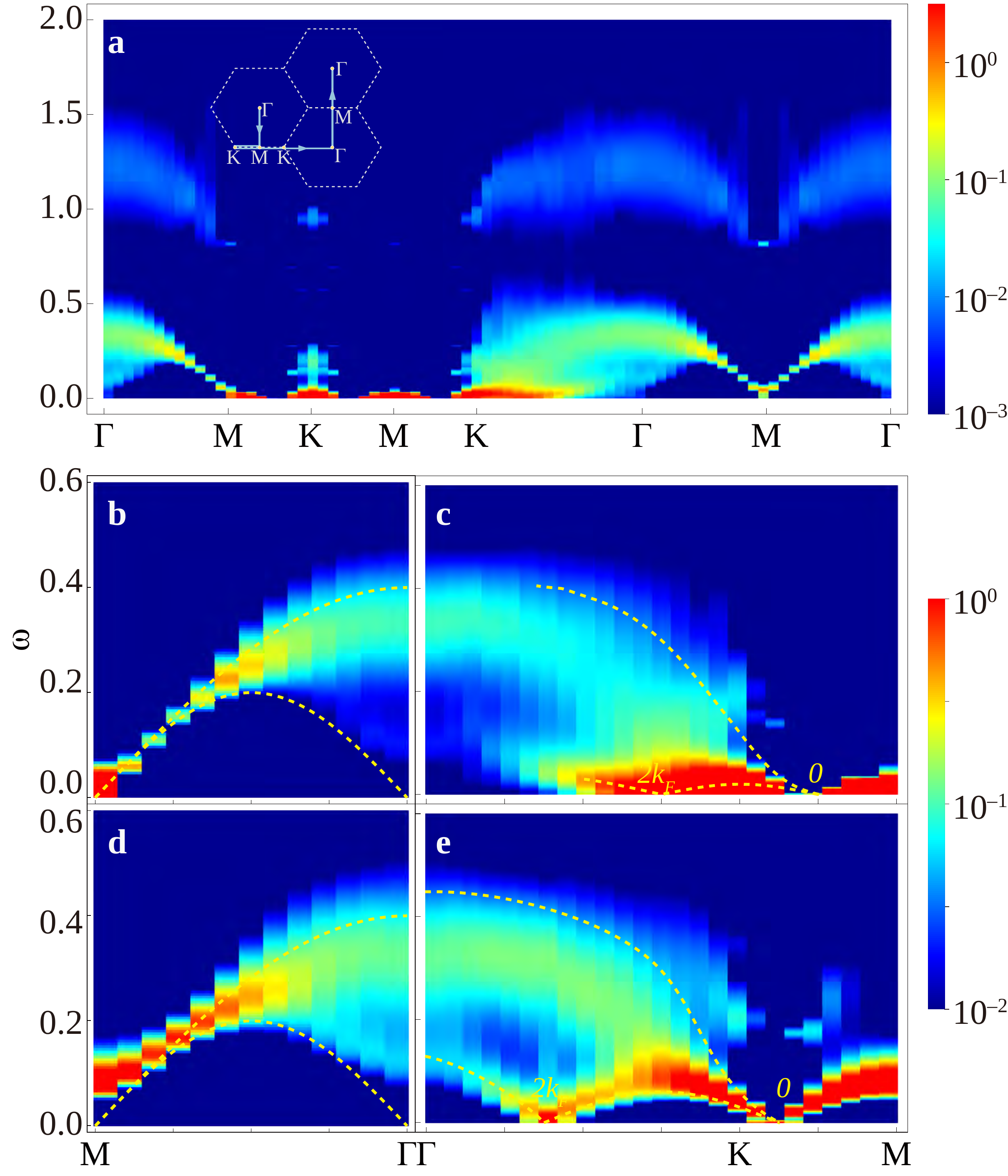}
    \caption{\textbf{The spectra at low string density.} \textbf{a}, The spectra at string density $\rho_\mathrm{s}=1/3$ measured at $\lambda=0.895$, $\Delta=0.2$, system size $L=24$ and temperature $T=1/96$ in the unit of $J$. The path taken in the Brillouin zone is specified in the inset. \textbf{b,c}, The zoom-in of low energy region are dived into parallel to the strings~(\textbf{b}) along $\mathrm{\Gamma M}$ line and perpendicular to the strings~(\textbf{c}) along $\mathrm{\Gamma KM}$ line respectively. \textbf{d,e}, The low energy excitation spectra measured at $\lambda=0.91$ and $\rho_\mathrm{s}=1/2$ includes parallel to the strings~(\textbf{d}) along $\mathrm{\Gamma M}$ line and perpendicular to the strings~(\textbf{e}) along $\mathrm{\Gamma KM}$ line respectively. The dashed lines denote the anticipated results in spin-$1/2$ XY chain~(\textbf{b,d}) and Luttinger liquid~(\textbf{c,e}). The gray italic momentum labels `0' and `$2k_\mathrm{F}$' are the momenta measured from the Fermi point of the projected 1d system. }
    \label{fig_3}
\end{figure}

We take $J'=0$ and tune $0<\lambda<1$ to fine-tune the string density to a comparatively low lever $\langle\rho_\mathrm{s}\rangle=1/3$ and $1/2$.
The energy hierarchy in the spectrum persists (Fig.~\ref{fig_3}a). The spinon branch and the string branch stay at the energy scale of $J$ and $\Delta$, respectively. To look more carefully into the string excitations, we scrutinise the low energy spectrum along $\mathrm{\Gamma M}$ and $\mathrm{\Gamma KM}$ lines (Fig.~\ref{fig_3}b--e).

The spectrum along the $\mathrm{\Gamma M}$ line corresponds to the internal string vibration. In the case of nearly independent strings at $\rho_\mathrm{s}=1/3$ (Fig.~\ref{fig_3}b), the spectrum is continuous and enveloped by two dome-like curves, most extensive at $\Gamma$ point and shrinking when moving towards $\mathrm{M}$ point. We find these features in good agreement quantitatively with the continuous spectrum of spin-$1/2$ XY chain (Fig.~\ref{fig_3}b, grey dashed lines) (See Supplementary Note 5), where $\Gamma$ and $\mathrm{M}$ points correspond to $k=\pi$ and $0$ momenta of the effective XY chain, respectively. On the other hand, when increasing the string density, the effective repulsion affects the internal vibration of strings, rendering the excitation gapped (Fig.~\ref{fig_3}d).

The spectrum along the $\Gamma\mathrm{KM}$ line corresponds to the dynamics that come from string interaction. The spectrum exhibits continuous features. Two nearly gapless points locate at $k_x=\pi(2-\langle\rho_\mathrm{s}\rangle)$ and $\pi(2-3\langle\rho_\mathrm{s}\rangle)$ (Fig.~\ref{fig_3}c,e). Whether these points are truly gapless depends on the commensuracy of the string density. At commensurate string density (such as $\langle\rho_\mathrm{s}\rangle=2/3$ in clock phase), the high order scattering process will open a gap in the excitation. However, such gap lies at a much lower energy level than our string description. The dispersions in the vicinity of the nearly gapless points are linear. These features hint at the spectroscopic similarity to Luttinger liquid~\cite{luttinger}~(See Supplementary Note 3). In fact, projecting onto the perpendicular direction, the strings can be described with spinless fermions with long-range repulsion moving in 1d. The two nearly gapless points correspond to the momentum points $0$ and $2k_\mathrm{F}$ measured from the Fermi point, with the Fermi momentum of the strings $k_\mathrm{F}=\pi\langle\rho_\mathrm{s}\rangle$, in accordance with the spectroscopic features of `Luttinger string liquid' (Fig.~\ref{fig_3}c,e, dashed lines).

\subsection{String and spinon spectra}

\begin{figure}[t!]
    \centering
    \includegraphics[width=.7\linewidth]{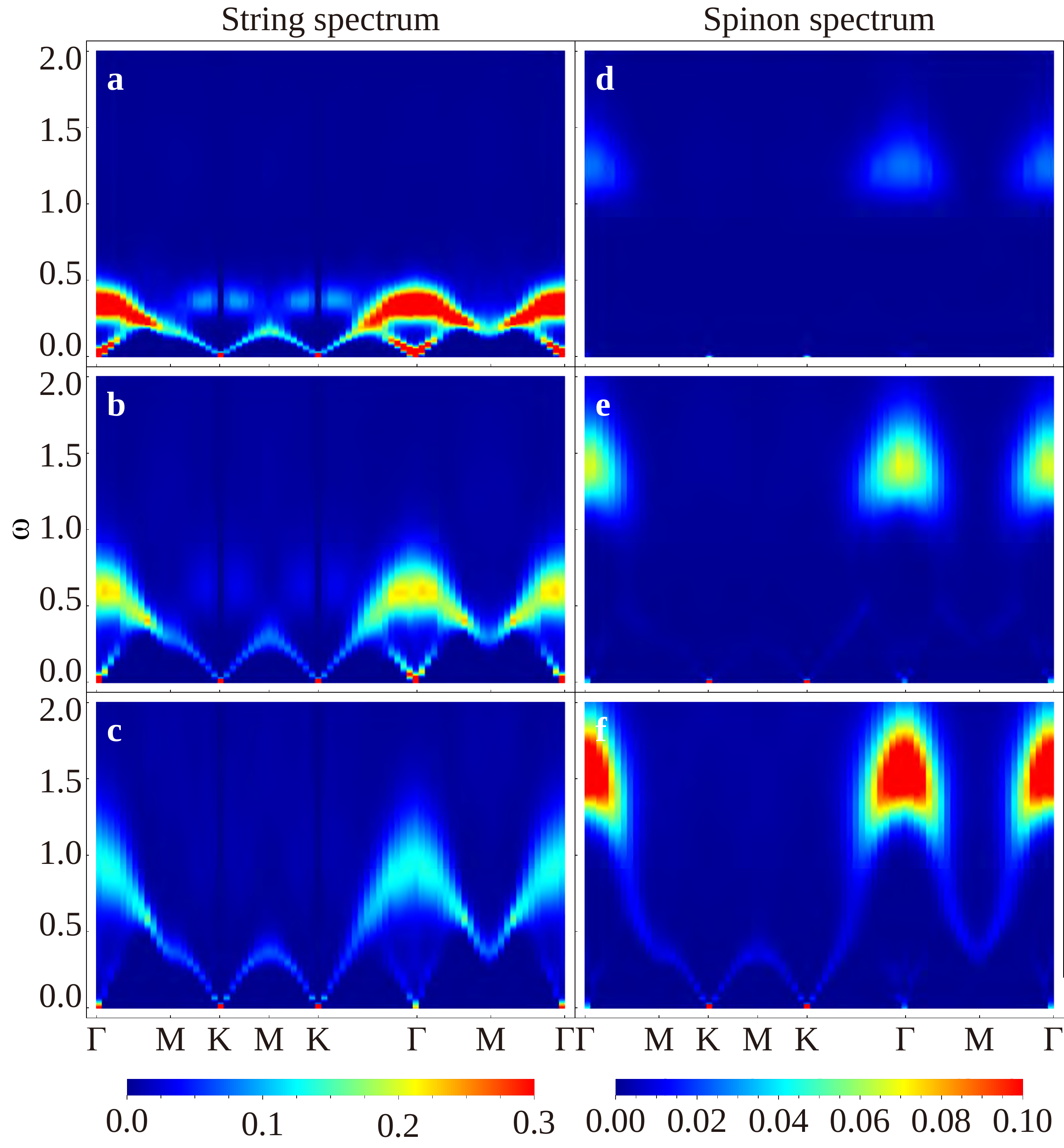}
    \caption{\textbf{The separated spectra of string excitations and spinon excitations.} String spectra~(\textbf{a--c}) and spinon spectra~(\textbf{d--f}) with $L=24$ and $T=1/96$ at $\Delta=0.20$, $J'=0.020$~(\textbf{a,d}), $\Delta=0.35$, $J'=0.035$~(\textbf{b,e}) and $\Delta=0.050$, $J'=0.050$~(\textbf{c,f}).}
    \label{fig_4}
\end{figure}

Having figured out the relation between different string-related excitations and their features in the spectrum, we now return to the isotropic case~(i.e., the exchange anisotropy is tuned to $\lambda=1$). To locate the excitations of different topological defects in the spectrum, we try to isolate the contributions from string and spinon excitations. We define a kink operator, which singles out the flippable spins at the kinks of the strings~(See Supplementary Note 4), and a spinon operator, which singles out the spins on triangles with the same spin polarization. As the kink operator is only sensitive to string excitations, and the spinon operator only to spinon excitations, their dynamic correlations isolate the contribution of two respective excitations -- At low field $\Delta=0.2$~(Fig.~\ref{fig_4}a,d), the string spectrum has only low energy part and the spinon spectrum have only high energy part; while for higher $\Delta$~(Fig.~\ref{fig_4}c,f), the energy scales of two excitations nearly coincide, and they mix up into what we see in the spin spectrum measured by neutron scattering.

\section{Discussion}

Our work focuses on the dynamical signature of quantum string excitations. We simulate the spectrum of frustrated TFIM on the triangular lattice, which is relevant to a large class of rare-earth magnets including the recently discovered TMGO. At low transverse field, we find continuum excitations at low energies. We show that the features of this continuum are in good agreement with the internal vibration and mutual interactions of quantum strings. In particular, the internal vibration (spectrum along $\Gamma\mathrm M$ line) matches well with the continuous spectrum of spin-$1/2$ XY chain, and the string interaction (spectrum along $\Gamma\mathrm{KM}$ line) induces excitations which exhibit the features of Luttinger liquid. 

Several trials can be considered in the experiment to observe the string-related excitations in the excitation spectra. If we can find materials with appropriate $\Delta/J$, we expect well separation of string-related excitations from spinons in excitation spectrum \cite{sl_expr_3,tmgo_kt,tmgo_neu,liao2021phase}.
In experiments, we can also study anisotropic models by applying in-plane uniaxial pressure. On the one hand, for non-Kramers doublet systems, the crystal field splitting $\Delta$ is generated and can be tuned by the applied pressure. This enables discovering string related excitations with appropriate $\Delta$. On the other hand, the applied pressure also produces exchange anisotropy $\lambda$, which tunes the string density $\rho_\mathrm{s}$. With sufficiently large $\lambda$ one would access the dilute string region where the intra-string excitations may be observed.
Besides rare-earth magnets, similar phenomena are also expected in other systems such as antiferroelectric materials~\cite{af_elec}, cold atoms~\cite{trap_ion} and quantum computer~\cite{QC}.

Our work not only provides insight into understanding the dynamical properties of triangular quantum Ising antiferromagnet such as TMGO, but also provides iconic features in the spectrum for the experimental search for emergent strings, as well as deepening the understanding of emergent quantum strings in general.

\section*{Methods}

\subsection{Stochastic series expansion~(SSE)}

We use SSE algorithm for the quantum Monte Carlo simulation of the $J_1$-$J_2$-$J_3$ Ising model~\cite{qmc_1,qmc_2,qmc_3}. In SSE, the evaluation of partition function $Z$ is done by a Taylor expansion, and the trace is taken by summing over a complete set of suitably chosen basis.
\begin{equation}
    Z=\sum_\alpha\sum_{n=0}^\infty\frac{\beta^n}{n!}\langle\alpha|(-H)^n|\alpha\rangle
\end{equation}
By writing the Hamiltonian as the sum of a set of operators whose matrix elements are easy to calculate $H=-\sum_iH_i$ and truncating the Taylor expansion at a sufficiently large cutoff $M$, we can further obtain
\begin{equation}
    Z=\sum_\alpha\sum_{\{i_p\}}\beta^n\frac{(M-n)!}{M!}\langle\alpha|\prod_{p=1}^nH_{i_p}|\alpha\rangle
\end{equation}
To carry out the summation, a Markov chain Monte Carlo procedure can be used to sample the operator sequence $\{i_p\}$ and the trial state $\alpha$. Diagonal update, where diagonal operators are inserted into and removed from the operator sequence, and cluster update, where diagonal and off-diagonal operates convert into each other, are adopted in the update strategy. For more details, see Supplementary Method 1.

\subsection{Stochastic analytical continuation~(SAC)}

We adopted the SAC method to obtain the spectral function $S(\omega)$ from the imaginary time correlation $G(\tau)$ measured from QMC~\cite{sac_1,sac_2,sac_3}. The spectral function $S(\omega)$ is connected to the imaginary time Green's function $G(\tau)$ through an integral equation $G(\tau)=\int_{-\infty}^{\infty}\mathrm{d}\omega S(\omega)K(\tau,\omega)$ where $K(\tau,\omega)$ is the kernal function . For spin systems, $K(\tau,\omega)=(e^{-\tau\omega}+e^{-(\beta-\tau)\omega})/\pi$. To ensure the normalization of spectral function, we further modify the transformation,
\begin{equation}
    G(\tau)=\int_0^\infty\frac{\mathrm{d}\omega}{\pi}\frac{e^{-\tau\omega}+e^{-(\beta-\tau)\omega}}{1+e^{-\beta\omega}}B(\omega)
    \label{eq:kernal}
\end{equation}
where $B(\omega)=S(\omega)(1+e^{-\beta\omega})$ is the renormalised spectral function.

In practice, $G(\tau)$ for a set of imaginary time $\tau_i(i=1,\dots N_\tau)$ is measured in QMC simulation together with the statistical errors. The renormalised spectral function is parametrised into large number of $\delta$-functions $B(\omega)=\sum_{i=0}^{N_\omega}a_i\delta(\omega-\omega_i)$ whose positions are sampled. Then the fitted Green's functions $\tilde{G}_i$ from Eq.~(\ref{eq:kernal}) and the measured Greens functions $\bar{G}_i$ are compared by the fitting goodness $\chi^2=\sum_{i,j=1}^{N_\tau}(\tilde{G}_i-\bar{G}_i)(C^{-1})_{ij}(\tilde{G}_j-\bar{G}_j)$ where $C_{ij}$ is the covariance matrix. A Metropolis process is utilised to update the series in sampling. The weight for a given spectrum is taken to follow a Boltzmann distribution $W(\{a_i,\omega_i\})\sim\exp\left(-\chi^2/2\Theta\right)$ with $\Theta$ a virtue temperature. All the sampled spectral functions are then averaged to obtain the final result. For more details, see the Supplementary Method 2. As an evidence of the validity of SAC, we also compare the spectral function obtained from SAC with the lowest excitation frequency extracted directly from the imaginary time correlations (see the Supplementary Figure 3).

\section*{Data availability}

The data that support the findings of this study are available from the authors upon reasonable request.

\section*{Code availability}

The computer codes that support the findings of this study are available from the authors upon reasonable request.

\section*{Acknowledgement}

We wish to thank Yang Zhou, Yang Qi, Zi-Yang Meng, Jun Zhao, Ying Jiang, and Yin-Chen He for the fruitful discussions. C. L. thanks Rong Yu for hospitality inviting him to visit Renmin University of China where part of the work is done. This work is supported by the National Key Research and Development Program of China (Grants Nos. 2017YFA0304204, 2016YFA0300501 and 2016YFA0300504), the National Natural Science Foundation of China (Grants Nos. 11625416 and 11474064), and the Shanghai Municipal Government (Grants Nos. 19XD1400700 and 19JC1412702). X.-F. Z. acknowledges funding from the National Science Foundation of China under Grants  No. 11874094 and No.12147102, Fundamental Research Funds for the Central Universities Grant No. 2021CDJZYJH-003. Z. Z. acknowledges support from the CURE (H.-C. Chin and T.-D. Lee Chinese Undergraduate Research Endowment) (19925) and National University Student Innovation Program (201910246148). C. L. is supported by the National Natural Science Foundation of China (11925402), Guangdong province (2016ZT06D348, 2020KCXTD001), the National Key R \& D Program (2016YFA0301700), Shenzhen High-level Special Fund (G02206304, G02206404), and the Science, Technology and Innovation Commission of Shenzhen Municipality (ZDSYS20170303165926217, JCYJ20170412152620376, KYTDPT20181011104202253), and Center for Computational Science and Engineering of SUSTech. The authors acknowledge Beijing PARATERA Tech CO.,Ltd.(\url{https://www.paratera.com/}) for providing HPC resources that have contributed to the research results reported within this paper.

\section*{Competing interests}

The authors declare no competing interests.

\section*{Author contributions}

X.-F. Z., Z. Z. and Z. Y. initiated the work. Z. Z. and Z. Y. performed the Monte Carlo simulations. All authors contributed to the analysis of the results. X.-F. Z. and Y. C. supervised the project.

\renewcommand{\figurename}{Supplementary Figure}
\setcounter{figure}{0}

\section{Supplementary Figure 1. Comparison with neutron spectrum}

\begin{figure}[t!]
    \includegraphics[width=0.8\linewidth]{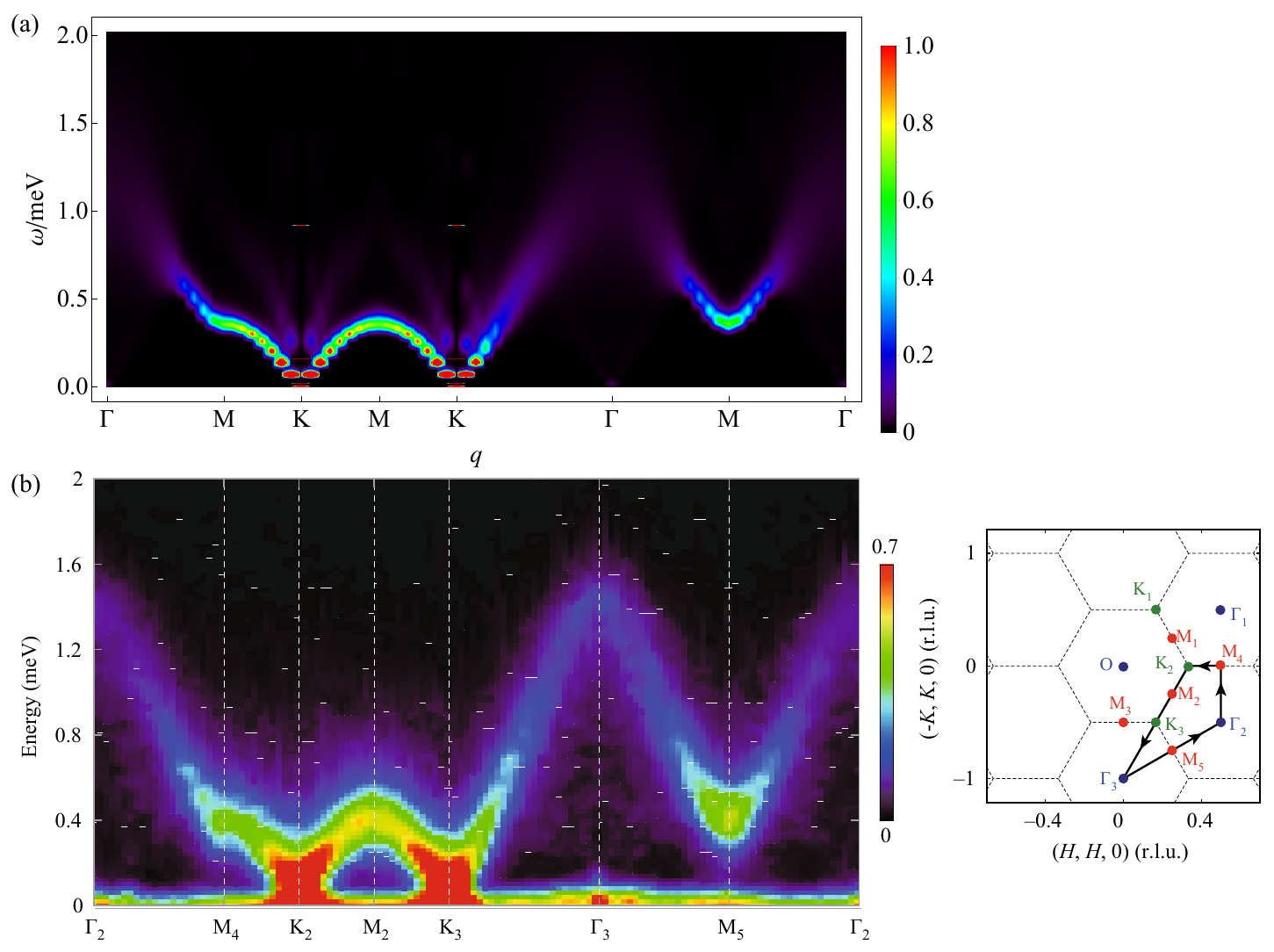}
    \caption{(a) The QMC-SAC simulation of the $S^{zz}$ spectrum, measured at $J=0.99\mathrm{meV}$, $J'=0.05\mathrm{meV}$ and $\Delta=0.54\mathrm{meV}$, $L=24$ and $\beta J=96$; (b) the inelastic neutron scattering result in $\mathrm{TmMgGaO}_4$, reproduced from Ref. \cite{sl_expr_3}. }
    \label{fig_s1}
\end{figure}

In Supplementary Figure~\ref{fig_s1}, we show that the QMC-SAC simulation on the antiferromagnetic Ising model with next-neareast-neighbour coupling and transverse field can reproduce the spectrum measured by inelastic neutron scattering \cite{sl_expr_3} on $\mathrm{TmMgGaO}_4$. As indicated by Ref. \cite{tmgo_kt}, the best fit parameters are found to be $J=0.99\mathrm{meV}$, $J'=0.05\mathrm{meV}$ and $\Delta=0.54\mathrm{meV}$. In the Fig. 1 of main text we have plotted the numerical results of the spectrum at such parameters in logarithm scale coloring so that the two modes can both be seen clearly. To compare this QMC-SAC spectrum with the neutron scattering result, we plotted the spectrum in linear scale coloring and take $\textrm{meV}$ as the energy unit. The spectrum from numerical simulation in Panel \textbf{a} and that from neutron scattering in Panel \textbf{b} agrees considerably well.

\section{Supplementary Figure 2. Spectra in linear scale}

\begin{figure}[t!]
    \centering
    \includegraphics[width=.6\linewidth]{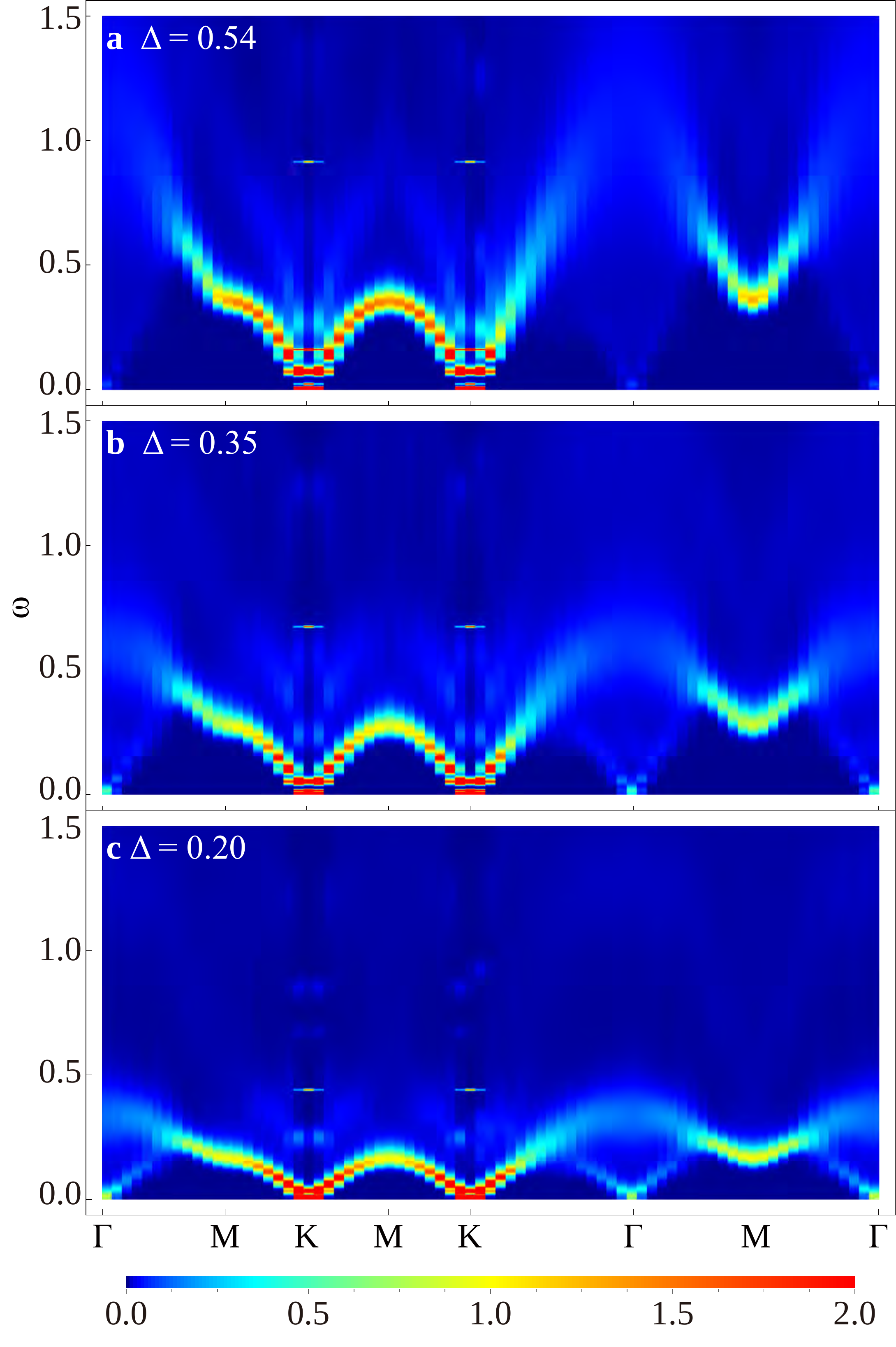}
    \caption{The spectra at different transverse field $\Delta$ (Fig.~1 in main text) plotted in linear scale at $\Delta=0.54$, $J'=0.05$ (\textbf{a}), $\Delta=0.35$, $J'=0.035$ (\textbf{b}) and $\Delta=0.20$, $J'=0.020$ (\textbf{c}). }
    \label{fig_s1_1}
\end{figure}

In Supplementary Figure~\ref{fig_s1_1}, we replot the the spectra at different transverse field $\Delta$ in Fig.~1 in the main text in linear scale.

\section{Supplementary Figure 3. Reliability check of stochastic analytical continuation}

\begin{figure}[t!]
    \centering
    \includegraphics[width=0.7\linewidth]{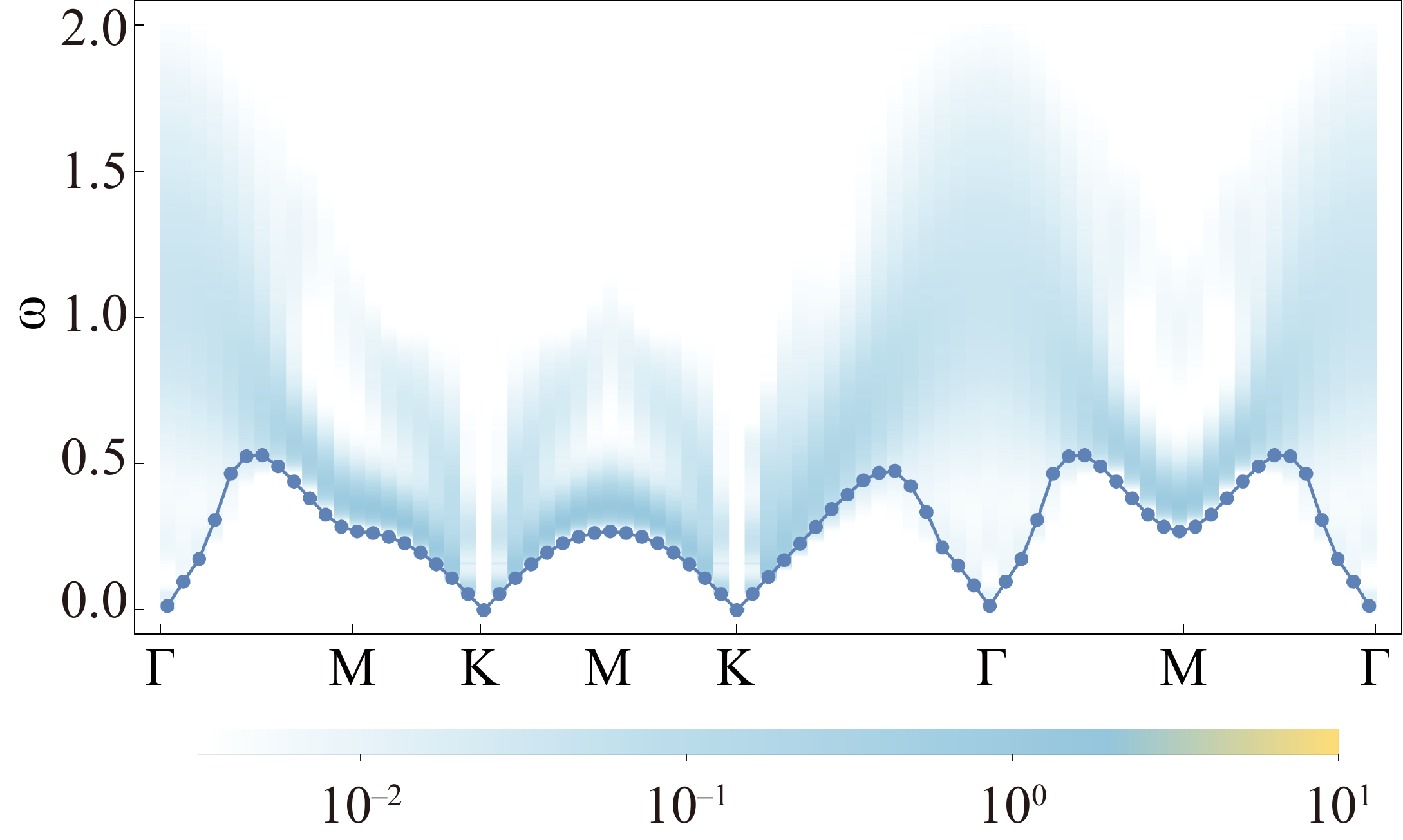}
    \caption{The spectrum calculated by the SAC, plotted together with lowest excitation frequency extracted directly from the imaginary time correlations. The data are measured at $\lambda=1,\Delta=0.54,J'=0.05$, same as Fig.~1a in the manuscript. }
    \label{fig_r1_1}
\end{figure}

In Supplementary Figure~\ref{fig_r1_1}, as an evidence of the validity of SAC, we also compare the spectral function obtained from SAC with the lowest excitation frequency extracted directly from the imaginary time correlations. As $G(\tau)\sim\int_0^\infty\mathrm{d}\tau\:S(\omega)\exp(-\omega\tau)$, at large $\tau$, only the lowest energy $\omega_m$ retains. Thus, we obtain $\omega_m$ by performing the fitting $G(\tau)\sim\exp(-\omega_m\tau)$ for $0.2\beta<\tau<0.4\beta$. The fitted $\omega_m$ agrees well with the low energy boundary of the SAC spectrum.

\section{Supplementary Figure 4. Visualisation of strings}

\begin{figure}[t!]
    \centering
    \includegraphics[width=.5\linewidth]{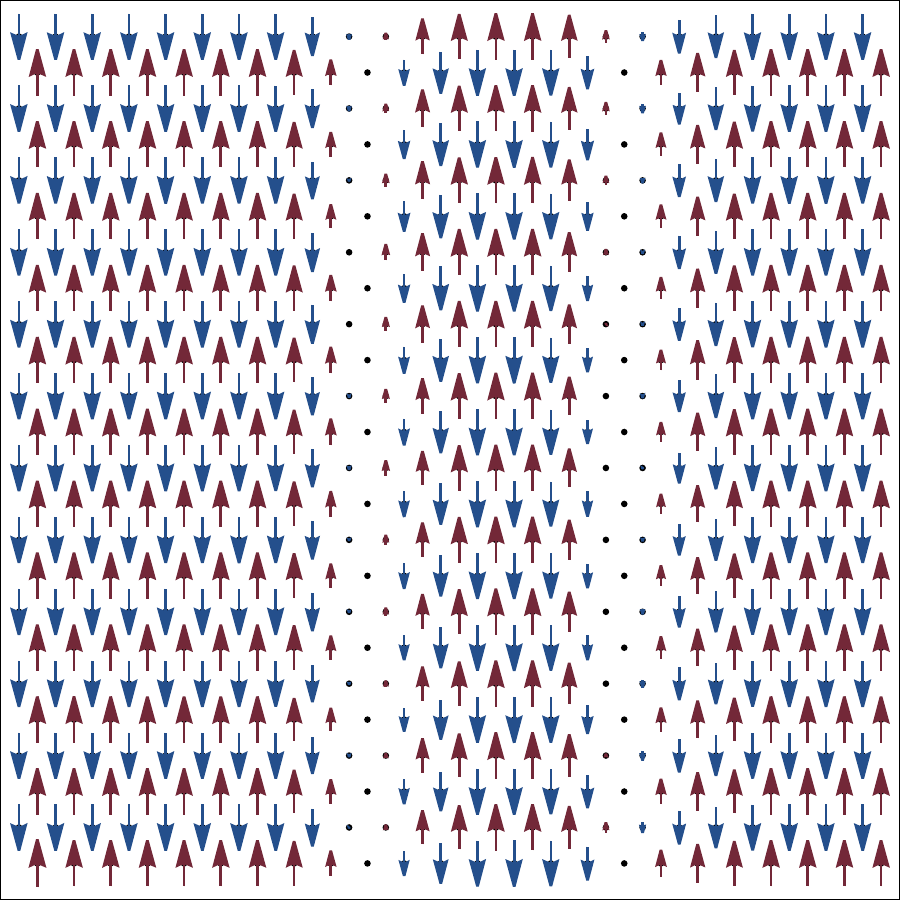}
    \caption{The snapshot of the real space spin correlation $G(\mathbf{r})=\langle S^z(\mathbf{r})S^z(0)\rangle$ measured at $\lambda=0.893,N_s=2$ in $L=24$ system. The strings fluctuate in the regions with weak spin correlation. }
    \label{fig_r1_2}
\end{figure}

It is possible to observe string-like configurations from Monte Carlo simulations. As an evidence, in Supplementary Figure~\ref{fig_r1_2} we measured a snapshot of the real-space spin correlation function $G(\mathbf{r})=\langle S^z(\mathbf{r})S^z(0)\rangle$ at $\lambda=0.893,N_s=2$ in $L=24$ system with rectangular periodic boundary within a short number of Monte Carlo steps, from which one can easily recognise the different stripe domains and the average string configuration as their domain walls. The average correlation near the strings are weak, revealing the strong fluctuation between two domains in this region.

\renewcommand{\figurename}{Figure for Supplementary Note}
\setcounter{figure}{0}

\section{Supplementary Note 1. Dimer and string mapping}

\begin{figure}[t!]
    \includegraphics[width=1\linewidth]{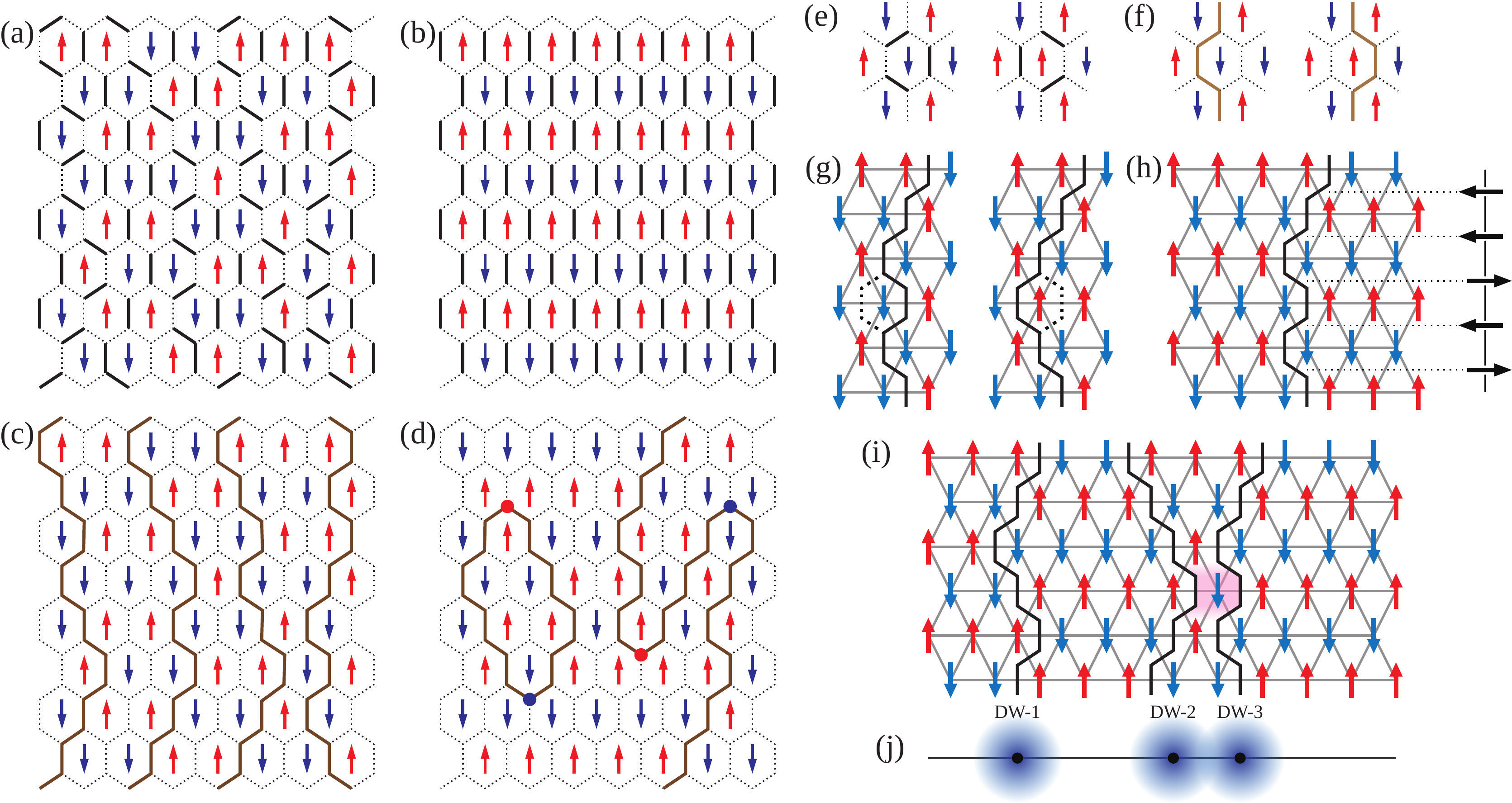}
    \caption{(a) Mapping of a spin configuration to a dimer covering on the dual hexagonal lattice; (b) the reference state of string mapping; (c) mapping a dimer covering to a string configuration and flippable spins are highlighted; (d) vortex and antivortex connected by two strings; (e) The dimer representation and (f) the string representation of flipping a flippable spin; (g) The deformation of the string by flipping spin; (h) The spin configuration with one string which can be mapped to spin-$1/2$ chain; (g) The spin configuration with three strings. At low temperature and small transverse field, the strings can not be crossed with each other (spin inside the red region can not flip due to triangle-rule), so that there exists effective interactions between strings; (h) The configuration in upper panel can be taken as three fermions with long-range effective interactions in one dimension.}
    \label{fig_s2}
\end{figure}

In this section we give a detailed description of the mapping between the spin configuration and the dimer configuration, and between the dimer configuration and the string configuration. The spin configuration is defined on triangular lattice. The dual lattice of this triangular lattice is a honeycomb lattice (Panel \textbf{a}). Each bond of the dual lattice sits on the perpendicular bisector of a bond of the original lattice. Each site of the original lattice is mapped onto a hexagonal plaquette of the dual lattice, and the six nearest-neighbour bonds connecting one site on the original lattice become the six edges of a hexagon on the dual lattice. Reciprocally, the triangles on the original lattice become the sites on the dual lattice, and the three bonds along the triangle are mapped onto the bonds connecting one site on the dual lattice. 

On this dual triangular lattice we can obtain the dimer representation of the spin configuration, where the degree of freedom resides on the bonds, i.e., each bond is either occupied or unoccupied with a dimer. Each original bond connecting parallel original spins is mapped into a dual occupied dimer, whereas each original bond connecting antiparallel spins is mapped into a dual unoccupied dimer (Panel \textbf{a}). The antiferromagnetic Ising coupling imposes on the original lattice the triangle rule, which dictates that one triangle either have two spin-downs and one spin-up, or two spin-ups and one spin-down. This constraint is mapped to an one-dimer-per-site constraint on the dual lattice. As each triangle satisfying the triangle rule on the original lattice is composed of two antiparallel bonds and one parallel bonds, each site on the dual lattice is connected to two unoccupied bons and one occupied bond. 

In this dimer representation we can further interpret the dynamics brought by the transverse field. To satisfy the constraint, the transverse field can only act on the spins such that three of its nearest neighbours point up and the other three point down, at the same time spin-ups and spin-downs appear alternatively. Such spins corresponds to the hexagonal plaquettes with three dimers and the occupied and unoccupied bonds appear alternatively (Panel \textbf{e}). By acting the transverse field, the spin is flipped, which exchanges its neighbouring bonds connecting parallel spins with bonds connecting antiparallel spins. On the dual lattice, this process corresponds to turning the occupied bonds into unoccupied bonds and vice versa on a hexagonal plaquette, which is exactly the dynamics of the off-diagonal term in the quantum dimer model. N.b., to keep the constraint, the transverse field has to be a perturbation with respect to the Ising term. In this way, the transverse field antiferromagnetic Ising model on triangular lattice is mapped exactly to a quantum dimer model with only the off-diagonal term, i.e., at $t/V=0$.

The dimer configuration can be further mapped onto a string configuration, where each site is connected to not one, but an even number of occupied links, and the links form closed loops. To perform the mapping, a reference configuration must be taken. Here we take the staggered dimer configuration where all the bonds in one direction are occupied (Panel \textbf{b}). This state corresponds to the stripe state in the original spin configuration. We then collapse the dimer configuration with the reference configuration. Each bond which is covered exactly once by the dimer configuration and the reference configuration is mapped into an occupied bond in the string configuration, while the bonds covered twice or uncovered by the two configurations are mapped into unoccupied bonds (Panel \textbf{c}). In this way, the dimers in dimer configuration are joint into directed, non-crossing closed strings which crosses the boundary in periodic boundary condition. 

Another way to understand such string configuration is to view it as a topological defect or domain wall on the original spin configuration, with respect to the stripe phase. Each original bond in $x$-direction connecting antiparallel spins is crossed by a string defined hereinbefore, i.e., each string is a domain wall which separates distinct stripe domains. 

Therefore we can analysis the dynamics in the string language. the flippable spins situate in the corners of the strings (Panel \textbf{g}). By flipping the spin with transverse field, the related strings are deformed in the form of creation or annihilation of kinks. E.g., in Panel \textbf{f}, when flipping the spin, the string on its left deforms to its right. 

The deformation of a single string can be analysed by mapping it to an effective spin model. Each string can be mapped into a spin-$1/2$ chain in the way that a left-going or right-going string segment is mapped into a spin-up or a spin-down respectively (Panel \textbf{h}). In this way the corners of the string are mapped into the kinks on the effective spin chain which separate antiparallel effective spins. By flipping the original spin inside the string corner, the left-going and right-going segment on the string is exchanged, i.e., the neighbouring spin-up and spin-down is exchanged. This corresponds to an effective XY interaction on the effective spin chain. In this way we can write down the effective Hamiltonian for a single string
\begin{equation}
    H_\mathrm{str}=\frac{\Delta}{4}\sum_{\langle ij\rangle}(\sigma'^+_i\sigma'^-_j+\sigma'^-_i\sigma'^+_j)
\end{equation}
Here $\sigma'_{i,j}$ denotes the effective spins of the string and $\sigma'^\pm$ are the ladder operators. The $\langle ij\rangle$ denotes that the sum is taken over all the nearest neighbours. This model is well studied and is exactly solvable under Jordan-Wigner transformation. Its excitations are free fermions in the form of effective ``spinons'' on the effective spin chain and ``kinks'' in the spin representation on the original lattice. 

The discussion above only considers single string case. When multiple strings appear in one configuration at the same time, they have to conform with the non-crossing condition. This condition is equivalent with the triangle rule in spin representation and the one-dimer-per-site constraint in dimer representation. Such condition restricts the Hilbert space of the strings. Same as the spin-$1/2$ chain, the size of a single string's Hilbert space is $2^{L_y}$. However, when two strings get close (e.g., the string 2 and 3 in Panel \textbf{i}), certain configurations with crossing strings are forbidden. So that the Hilbert space shrinks and an effective long-range interaction emerges (Panel \textbf{i}). Such interaction is found repulsive and should be algebraic to the distance between strings \cite{incom_1}. On the contrary, violation of this rule results in the triangles with three parallel spins. Such triangle hosts a spinon excitation, which is also known as ``vortex'' as the winding number of effective $U(1)$ spin is non-zero \cite{tmgo_kt}. The vortices are connected by either a string loop or a non-directed string (Panel \textbf{d}).

\section{Supplementary Note 2. Incommensurate order}

\begin{figure}[t!]
    \centering
    \includegraphics[width=1\linewidth]{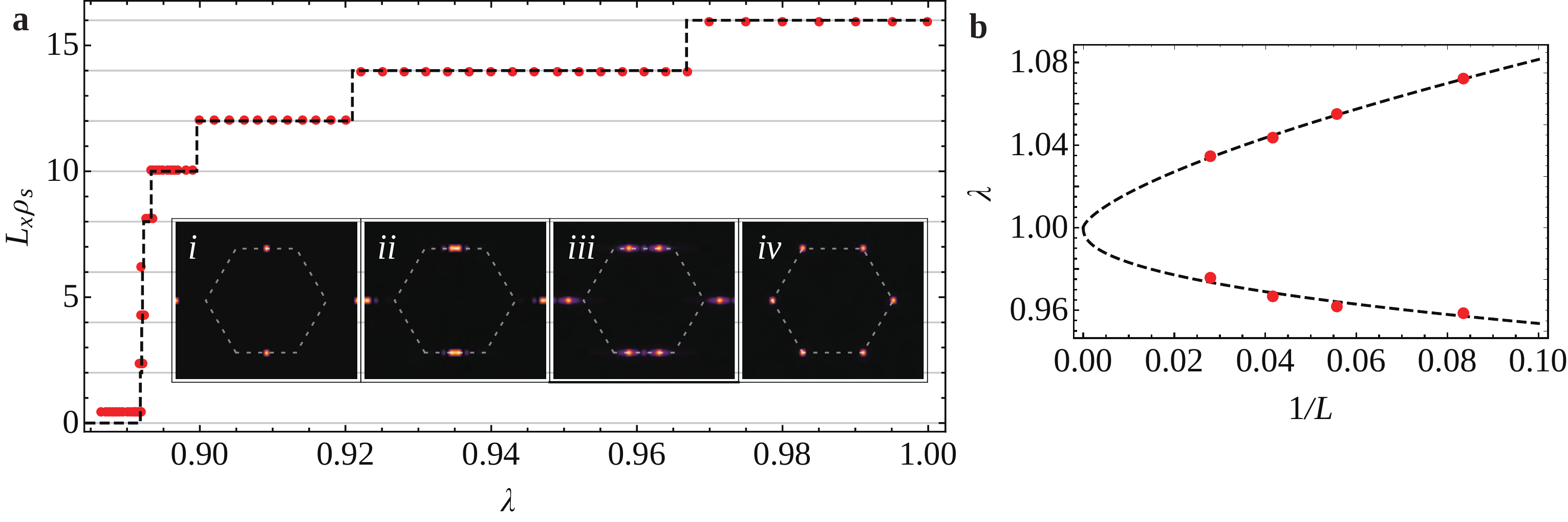}
    \caption{(a) the string density--anisotropy relation, measured at $\beta=L=24$ and $J'=0$; Inset : The magnetic structure factor with different string density, at (i) $\lambda=0.89$, $\rho_\mathrm{s}=0$ (stripe phase); (ii) $\lambda=0.892$, $\rho_\mathrm{s}=1/12$; (iii) $\lambda=0.895$, $\rho_\mathrm{s}=1/3$; and (iv) $\lambda=1.00$, $\rho_\mathrm{s}=2/3$; (b) The finite size scaling result for the upper and lower boundary of the clock order plateau $\rho_s=2/3$.}
    \label{fig_s2_1}
\end{figure}

By decreasing $\lambda$ from 1, the string density $\rho_\mathrm{s}$, defined as 
\begin{equation}
    \rho_s=\frac{1}{N}\sum_{\langle ij\rangle_x}\frac{1-4S_i^zS_j^z}{2}
\end{equation}
which effectively counts the number of horizontal bonds connecting opposite spins, drops from $2/3$ and finally reach zero (Panel \textbf{a}) as the system enters the stripe phase through a continuous phase transition. Accordingly, the peak of magnetic structure factor
\begin{equation}
    S(\mathbf{q})=\frac{1}{N}\sum_{ij}\langle S_i^zS_j^z\rangle\exp\left(\mathrm{i}\mathbf{q}\cdot(\mathbf{r}_i-\mathbf{r}_j)\right)
\end{equation}
moves from $\mathrm{K}$ to $\mathrm{M}$ linearly with the string density (Panel \textbf{a} inset), in accordance with the theoretical expectation of wave vector $\mathbf{q}=(\pm(2-\rho_s),0)\pi$ and $(\pm\rho_s,\pm 2/\sqrt{3})\pi$. 

It should also be noted that the structure of plateaux with discrete jumps is a result of the topological nature of the strings together with the finite size effect. Each of the plateau corresponding to an even integer number of domain walls $N_s=\rho_sL_x$. The platforms and jumps become increasingly continuous for larger system scale. A finite-size scaling (Panel \textbf{b}) of the left and right jumping point of the $\rho_s=2/3$ platform corresponding to the clock phase demonstrate that in the thermodynamic limit the domain wall density should be a continuous function of $J_x$.

\section{Supplementary Note 3. Resemblance to the Luttinger liquid}

\begin{figure}[t!]
    \centering
    \includegraphics[width=1\linewidth]{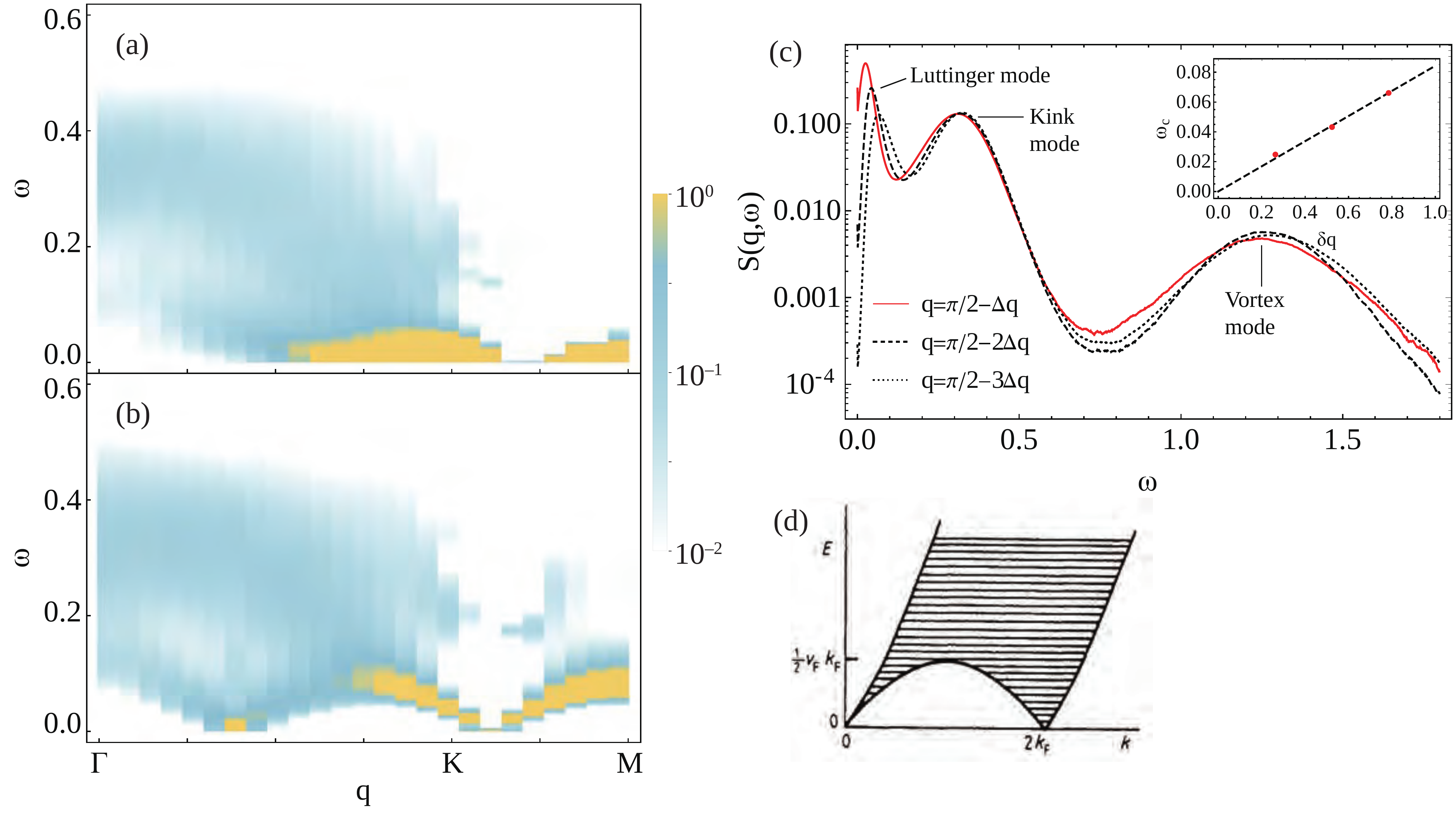}
    \caption{The low frequency region of $S^{zz}$ spectra along $x$-direction measured at (a) $\rho_\mathrm{s}=1/3$, $J_x=0.895$ and (b) $\rho_\mathrm{s}=1/2$, $J_x=0.91$, the dashed lines show the correspondence to Luttinger liquid; (c) the spectral function at momenta in vicinity of $q=\pi/2$; inset: the center of Luttinger mode as a function of momentum deviation at $\rho_\mathrm{s}=1/2$, $J_x=0.91$. $L_x=L_y=24$ and $\beta=96$ is taken, $\Delta q=2\pi/L_x$ is the minimal momentum interval; (d) The particle-hole spectrum of Luttinger liquid, reproduced from Ref. \cite{haldane}. }
    \label{fig_s3}
\end{figure}

In this section, we demonstrate that the features in the $S^{zz}$ spectrum at low string density $\rho_s$ are reminiscent of the Luttinger liquid when we add an anisotropy As described previously, in the perpendicular direction, the strings can be viewed as 0D fermions moving on 1D line. There exists a long-range power-law repulsive interaction between adjacent fermions. As indicated in Ref. \cite{fermi_lutt}, this may result in a Luttinger liquid phase at certain parameter region. 

The spectrum of an one dimensional Luttinger liquid is characterized by two gapless points at $q=0$ and $q=2k_F$, measured from the Fermi surface, where $k_F$ denotes the Fermi wave vector. The $k_F$ is a function of the string density $\rho_s$ (or particle density in the 1D)
\begin{equation}
    k_F=\pi\rho_s
\end{equation}
In the $S^{zz}$ spectrum, the $q=0$ point is mapped to $\mathbf{q}=((2-\rho_s)\pi,0)$, and the $q=2k_F$ point is mapped to $\mathbf{q}=((2-3\rho_s)\pi,0)$. The dispersion in vicinity of these gapless points are linear. These two gapless points and the linear dispersion in vicinity are indeed observed in numerical results (Panel \textbf{a}, \textbf{b}). 

We then scrutinize the spectra at momenta in vicinity of the $q=2k_F$ point (Panel \textbf{c}). Three distinct peak is observed. The lowest mode owns a frequency proportional to the momentum deviation, thus corresponds to the excitations of Luttinger liquid. The medium mode situates near the frequency of transverse field $\Gamma$, thus corresponds to the kink excitation (or the effective spinon excitation in the effective XY model). These two peaks together compose the excitations related to quantum strings. The highest mode corresponds to the triangle-rule breaking vortex excitation at the energy scale of $J$.

\section{Supplementary Note 4. Definition of kink and vortex operator}

\begin{figure}[t!]
    \centering
    \includegraphics[width=0.8\linewidth]{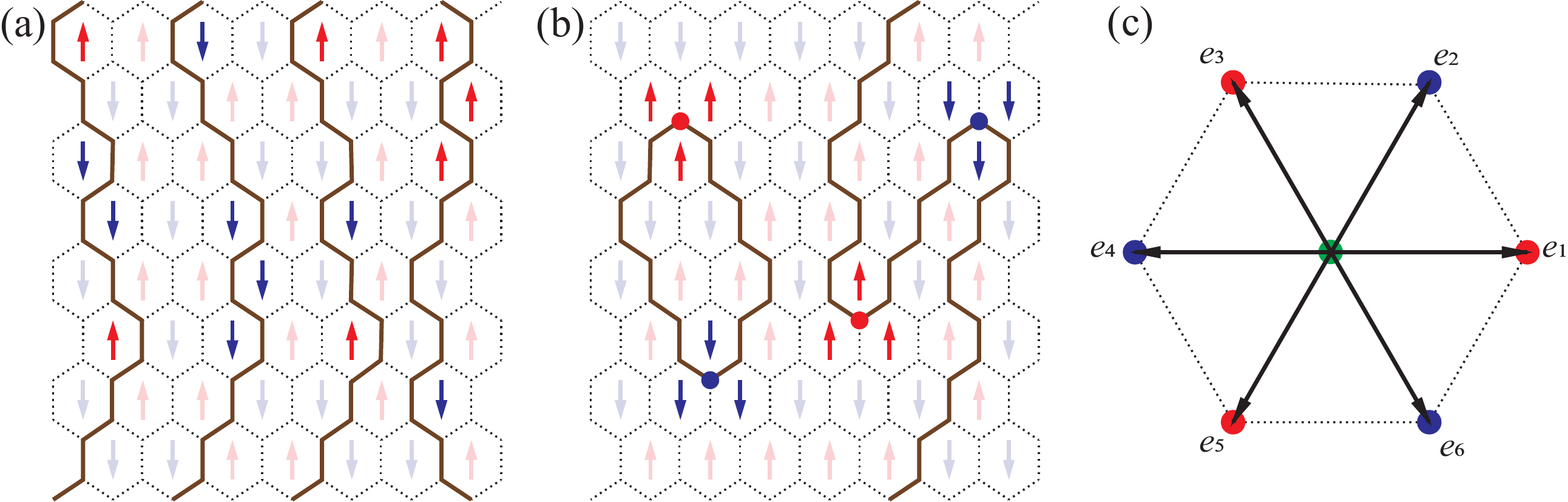}
    \caption{Demonstration of the kink and vortex operator. The red up arrows and blue down arrows show where (a) kink operator and (b) vortex operator is non-zero, whereas the whitened arrows show where the operators vanish.}
    \label{fig_s4}
\end{figure}

To isolate the contributions from string and vortex mode, we define the kink and vortex operator. The spectra calculated from their correlation can effectively distinguish the strength of $S^{zz}$ spectrum which comes from string and vortex mode. These two operators single out the spins concerned with corresponding excitation. 

For the string mode we define a kink operator. As mentioned before, the dynamics of the strings can be mapped into an effective spin-$1/2$ chain with XY nearest neighbour coupling. The excitation of such chain is fermions known as kinks residing on each bond connecting opposite effective spins. This corresponds to the corners of strings in the 2D string representation. In the spin representation, the spin envelopped by the corner has three spin-ups and three spin-downs as its neighbours which appear alternatively. This can be used as a criteria to single out the kinks. The kink operator is defined on each site of the spin configuration. It is non-zero only on the sites which is envelopped by a corner of the string, or to written explicitly, 
\begin{equation}
    \mathscr{K}(\mathbf{r})=\left\{\begin{array}{ll}
        S(\mathbf{r}),&\textrm{if }S(\mathbf{r}+\mathbf{e}_1)=-S(\mathbf{r}+\mathbf{e}_2)=S(\mathbf{r}+\mathbf{e}_3)\\
        &\quad=-S(\mathbf{r}+\mathbf{e}_4)=S(\mathbf{r}+\mathbf{e}_5)=-S(\mathbf{r}+\mathbf{e}_6)\\
        0,&\textrm{elsewhere.}
    \end{array}\right.
\end{equation}
where $\mathbf{e}_i$ is the displacement from a site to its six neighbours (Panel \textbf{c}). N.b., this definition is equivalent to singling out the ``flippable'' spins.

On the other hand, for the vortex mode we define a vortex operator. In the original spin representation, the vortices take the form of a triangle with three parallel spins. So we just have to single out the spins which belong to triangles with parallel spins, or to written explicitly
\begin{equation}
    \mathscr{V}(\mathbf{r})=\left\{\begin{array}{ll}
        S(\mathbf{r}),&\textrm{if }S(\mathbf{r})=S(\mathbf{r}+\mathbf{e}_1)=S(\mathbf{r}+\mathbf{e}_2)\textrm{ or }S(\mathbf{r})=S(\mathbf{r}+\mathbf{e}_2)=S(\mathbf{r}+\mathbf{e}_3)\\
        &\textrm{or }S(\mathbf{r})=S(\mathbf{r}+\mathbf{e}_3)=S(\mathbf{r}+\mathbf{e}_4)\textrm{ or }S(\mathbf{r})=S(\mathbf{r}+\mathbf{e}_4)=S(\mathbf{r}+\mathbf{e}_5)\\
        &\textrm{or }S(\mathbf{r})=S(\mathbf{r}+\mathbf{e}_5)=S(\mathbf{r}+\mathbf{e}_6)\textrm{ or }S(\mathbf{r})=S(\mathbf{r}+\mathbf{e}_6)=S(\mathbf{r}+\mathbf{e}_1)\\
        0,&\textrm{elsewhere.}
    \end{array}\right.
\end{equation}

\section{Supplementary Note 5. Comparison with spin wave methods}

\begin{figure}[t!]
    \centering
    \includegraphics[width=0.48\linewidth]{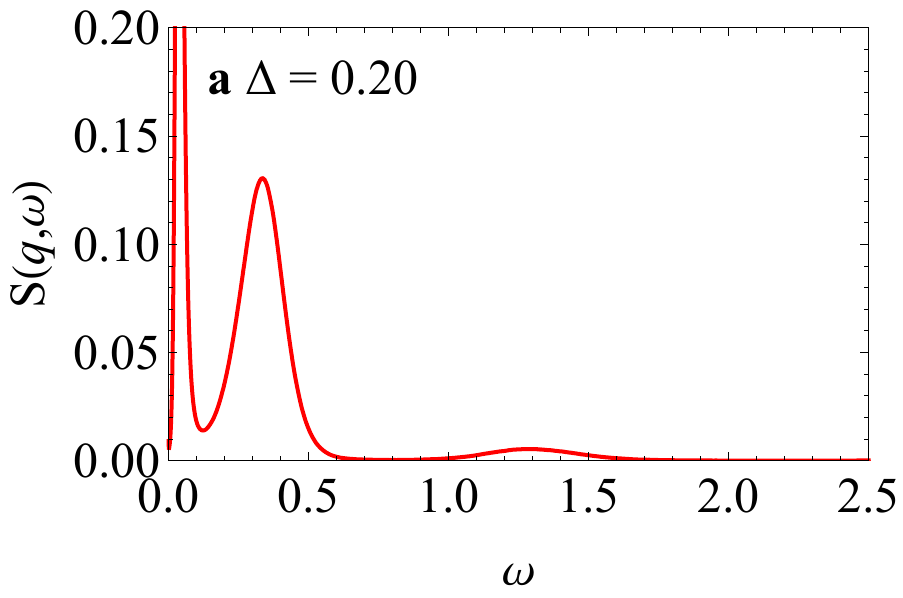}
    \caption{The spectral function $S(\mathbf{q},\omega)$ at momentum $q_x=\Delta q=2\pi/L,q_y=0$ measured at $\Gamma=0.2$, plotted in linear scale} 
    \label{fig_r2_1}
\end{figure}

In this section, we compare our string description with the spin wave descriptions that have been formerly used to study the spectrum of rare-earth compounds. 

First, the spin wave theories are semi-quantitative theories. Particularly, in frustrated magnets, quantum fluctuation is largely neglected in spin waves. As a matter of fact, the predicted parameters, especially the transverse field $\Delta$, for TMGO by spin wave theory deviates a lot from those extracted by more methods such as QMC. Therefore, tuning parameters away from their physical value is needed to produce spectra that agree with numerics or experiments. 

By contrast, our string picture can produce quite simple and effective quantitative predictions in a wide range of system parameters. The most representative example is the envelop of continuum excitations. Along $\Gamma\mathrm{M}$ direction, we predict that the excitations correspond to kinks inside each string, and the spectra are enveloped by the sine curves $\omega_1=\Delta\cos\sqrt{3}k_y/2$ and $2\Delta\cos\sqrt{3}k_y/4$, which agrees considerably well with numerical results (Fig. 3b in main text). Along $\Gamma\mathrm{KM}$ direction, we predict the excitations correspond to Luttinger liquids above the energy scale of the gap. The nearly gapless point should lie at $k_x=2\pi-\pi\rho_s$ and $k_x=2\pi-\pi\rho_s$, and the dispersions in vicinity should be linear, which also agree with our numerical results. As far as we know, our string picture is the simplest theory that explains these features altogether. 

Meanwhile, some features of our spectrum are hard to be reproduced by spin wave theories. For example, at low transverse field $\Delta=0.2$, in vicinity of $\Gamma$ point, we find three instinct peaks in the spectrum, centered at $\omega=0.038,0.337$ and $1.302$ respectively, composing $37\%,58\%$ and $5\%$ of the spectrum weight. These three peaks are explained in our string picture as the excitation due to interstring interaction, kink-antikink pair on single string and the spinons, respectively. On the other hand, spin wave theory only predicts the existence of two peaks --- two of the three modes in the linear spin wave are degenerate at $\Gamma$ point required by degeneracy, as shown in Fig.~9 of Ref.~\cite{mat_lgt_1}. The two modes predicted by LSW corresponds to the highest frequency and lowest frequency peak in the spectrum we find. The peak that makes up the largest portion of weight is missed in the linear spin wave theory. This peak is unlikely to come from the non-linear corrections as well, since its weight is larger than the LSW contribution. 

Our string picture becomes more rigorous and provide richer physics when we consider anisotropic NN interactions. In this case, the spatial anisotropy can stabilize the incommensurate phase out of the commensurate clock phase. The incommensurate order wave vector exhibits a linear relation with the string density. In the string picture, by minimizing the string energy, we are able to predict a string density-anisotropy relation that agrees with the numerical results (See Supplementary Note 1, 2). We are also able to clearly visualise the strings in the spin configuration snapshot~(See Supplementary Figure 4). Furthermore, at low string density with weak inter-string interactions, the spectra exhibit clearer continuous features with agree perfectly with the effective model.

\renewcommand{\figurename}{Figure for Supplementary Methods}
\setcounter{figure}{0}

\renewcommand{\thefigure}{\!}

\section{Supplementary Method 1. Stochastic series expansion (SSE)}

For the numerical works in this paper, we use a quantum Monte Carlo (QMC) method with stochastic series expansion (SSE) algorithm\cite{qmc_1,qmc_2,qmc_3} to calculate the ground state properties and imaginary time Green function. This method will be briefly introduced below. 

In quantum statistics, the measurement of observables is closely related to the calculation of partition function $Z$
\begin{equation}
    \langle\mathscr{O}\rangle=\mathrm{tr}\:\left(\mathscr{O}\exp(-\beta H)\right)/Z,\quad Z=\mathrm{tr}\exp(-\beta H)
\end{equation}
where $\beta=1/T$ is the inverse temperature, $H$ is the Hamiltonian of the system and $\mathscr{O}$ is an arbitrary observable. Typically, in order to evaluate the ground state property, one takes a sufficiently large $\beta$ such that $\beta\sim L^z$, where $L$ is the system scale and $z$ is the dynamical exponent. In SSE, such evaluation of $Z$ is done by a Taylor expansion of the exponential and the trace is taken by summing over a complete set of suitably-chosen basis. 
\begin{equation}
    Z=\sum_\alpha\sum_{n=0}^\infty\frac{\beta^n}{n!}\langle\alpha|(-H)^n|\alpha\rangle
\end{equation}
We then write the Hamiltonian as the sum of a set of operators whose matrix elements are easy to calculate. 
\begin{equation}
    H=-\sum_iH_i
\end{equation}
In practice we truncate the Taylor expansion at a sufficiently large cutoff $M$ and it is convenient to fix the sequence length by introducing in identity operator $H_0=1$ to fill in all the empty positions despite it is not part of the Hamiltonian. 
\begin{equation}
    (-H)^n=\sum_{\{i_p\}}\prod_{p=1}^nH_{i_p}=\sum_{\{i_p\}}\frac{(M-n)!n!}{M!}\prod_{p=1}^nH_{i_p}
\end{equation}
and 
\begin{equation}
    Z=\sum_\alpha\sum_{\{i_p\}}\beta^n\frac{(M-n)!}{M!}\langle\alpha|\prod_{p=1}^nH_{i_p}|\alpha\rangle
\end{equation}

To the carry out the summation, a Monte Carlo procedure can be used to sample the operator sequence $\{i_p\}$ and the trial state $\alpha$ with according to their relative weight
\begin{equation}
    W(\alpha,\{i_p\})=\beta^n\frac{(M-n)!}{M!}\langle\alpha|\prod_{p=1}^nH_{i_p}|\alpha\rangle
\end{equation}
For sampling we adopt a Metropolis algorithm where the configuration of one step is generated based on updating the configuration of the former step and the update is accepted at a probability
\begin{equation}
    P(\alpha,\{i_p\}\rightarrow\alpha',\{i'_p\})=\min\left(1,\frac{W(\alpha',\{i'_p\})}{W(\alpha,\{i_p\})}\right)
\end{equation}
Diagonal update, where diagonal operators are inserted into and removed from the operator sequence, and cluster update, where diagonal and off-diagonal operates convert into each other, are adopted in update strategy. 

In transverse field Ising model $H=J\sum_bS_{i_b}^zS_{j_b}^z-h\sum_i\sigma_i^x$, we write the Hamiltonian as the sum of following operators
\begin{equation}
    \begin{aligned}
        H_0&=1\\
        H_i&=h(S_i^++S_i^-)/2\\
        H_{i+n}&=h/2\\
        H_{b+2n}&=J(1/4-S_{i_b}^zS_{j_b}^z)
    \end{aligned}
\end{equation}
where a constant is added into the Hamiltonian for convenience. For the non-local update, a branching cluster update strategy is constructed \cite{qmc_2}, where a cluster is formed in $(D+1)$-dimensional by grouping spins and operators together. Each cluster terminates on site operators and includes bond operators (Panel \textbf{a}). All the spins in each cluster is flipped together at a probability $1/2$ after all clusters are identified.

\section{Supplementary Method 2. Stochastic analytical continuation (SAC)}

For the spectra in this paper we adopted a stochastic analytical continuation (SAC) \cite{sac_1,sac_2,sac_3} method to obtain the spectral function $S(\omega)$ from the imaginary time correlation $G(\tau)$ measured from QMC, which is generally believed a numerically unstable problem. This method will be briefly introduced below. 

The spectral function $S(\omega)$ is connected to the imaginary time Green's function $G(\tau)$ through an integral equation 
\begin{equation}
    G(\tau)=\int_{-\infty}^{\infty}\mathrm{d}\omega S(\omega)K(\tau,\omega)
\end{equation}
where $K(\tau,\omega)$ is the kernel function depending on the temperature and the statistics of the particles. We restrict ourselves to the case of spin systems and with only positive frequencies in the spectral, where $K(\tau,\omega)=(e^{-\tau\omega}+e^{-(\beta-\tau)\omega})/\pi$. To ensure the normalization of spectral function, we further modify the transformation and come to the following equation :
\begin{equation}
    G(\tau)=\int_0^\infty\frac{\mathrm{d}\omega}{\pi}\frac{e^{-\tau\omega}+e^{-(\beta-\tau)\omega}}{1+e^{-\beta\omega}}B(\omega)
    \label{eq_kernal}
\end{equation}
where $B(\omega)=S(\omega)(1+e^{-\beta\omega})$ is the renormalized spectral function. 

In practice, $G(\tau)$ for a set of imaginary time $\tau_i(i=1,\cdots N_\tau)$ is measured in QMC simulation together with the statistical errors. The renormalized spectral function is parametrized into large number of equal-amplitude $\delta$-functions whose positions are sampled (\textbf{b})
\begin{equation}
    B(\omega)=\sum_{i=0}^{N_\omega}a_i\delta(\omega-\omega_i)
\end{equation}
Then the fitted Green's functions $\tilde{G}_i$ from Eq. \ref{eq_kernal} and the measured Greens functions $\bar{G}_i$ are compared by the fitting goodness
\begin{equation}
    \chi^2=\sum_{i,j=1}^{N_\tau}(\tilde{G}_i-\bar{G}_i)(C^{-1})_{ij}(\tilde{G}_j-\bar{G}_j)
\end{equation}
where the covariance matrix is defined as 
\begin{equation}
    C_{ij}=\frac{1}{N_B(N_B-1)}\sum_{b=1}^{N_B}(G_i^b-\bar{G}_i)(G_j^b-\bar{G}_j),
\end{equation}
with $N_B$ the number of bins, the measured Green's functions of each $G_i^b$. 

A Metropolis process is utilized to update the series in sampling. The weight for a given spectrum is taken to follow a Boltzmann distribution
\begin{equation}
    W(\{a_i,\omega_i\})\sim\exp\left(-\frac{\chi^2}{2\Theta}\right)
\end{equation}
with $\Theta$ a virtue temperature to balance the goodness of fitting $\chi^2$ and the smoothness of the spectral function. All the spectral functions of sampled series $\{a_i,\omega_i\}$ is then averaged to obtain the spectrum as the final result. 


\begin{thebibliography}{10}
    \expandafter\ifx\csname url\endcsname\relax
      \def\url#1{\texttt{#1}}\fi
    \expandafter\ifx\csname urlprefix\endcsname\relax\def\urlprefix{URL }\fi
    \providecommand{\bibinfo}[2]{#2}
    \providecommand{\eprint}[2][]{\url{#2}}
    
    \bibitem{frus_1}
    \bibinfo{author}{Moessner, R.} \& \bibinfo{author}{Ramirez, A.~P.}
    \newblock \bibinfo{title}{Geometrical frustration}.
    \newblock \emph{\bibinfo{journal}{Phys. Today}} \textbf{\bibinfo{volume}{59}},
      \bibinfo{pages}{24--29} (\bibinfo{year}{2006}).
    
    \bibitem{frus_2}
    \bibinfo{author}{Vojta, M.}
    \newblock \bibinfo{title}{Frustration and quantum criticality}.
    \newblock \emph{\bibinfo{journal}{Rep. Prog. Phys.}}
      \textbf{\bibinfo{volume}{81}}, \bibinfo{pages}{064501}
      (\bibinfo{year}{2018}).
    
    \bibitem{gauge_1}
    \bibinfo{author}{Read, N.} \& \bibinfo{author}{Sachdev, S.}
    \newblock \bibinfo{title}{Large-${N}$ expansion for frustrated quantum
      antiferromagnets}.
    \newblock \emph{\bibinfo{journal}{Phys. Rev. Lett.}}
      \textbf{\bibinfo{volume}{66}}, \bibinfo{pages}{1773--1776}
      (\bibinfo{year}{1991}).
    
    \bibitem{gauge_2}
    \bibinfo{author}{Isakov, S.~V.}, \bibinfo{author}{Gregor, K.},
      \bibinfo{author}{Moessner, R.} \& \bibinfo{author}{Sondhi, S.~L.}
    \newblock \bibinfo{title}{Dipolar spin correlations in classical pyrochlore
      magnets}.
    \newblock \emph{\bibinfo{journal}{Phys. Rev. Lett.}}
      \textbf{\bibinfo{volume}{93}}, \bibinfo{pages}{167204}
      (\bibinfo{year}{2004}).
    
    \bibitem{gauge_3}
    \bibinfo{author}{Kitaev, A.}
    \newblock \bibinfo{title}{Anyons in an exactly solved model and beyond}.
    \newblock \emph{\bibinfo{journal}{Ann. Phys. (N. Y.)}}
      \textbf{\bibinfo{volume}{321}}, \bibinfo{pages}{2--111}
      (\bibinfo{year}{2006}).
    
    \bibitem{gauge_4}
    \bibinfo{author}{Balents, L.}
    \newblock \bibinfo{title}{Spin liquids in frustrated magnets}.
    \newblock \emph{\bibinfo{journal}{Nature}} \textbf{\bibinfo{volume}{464}},
      \bibinfo{pages}{199--208} (\bibinfo{year}{2010}).
    
    \bibitem{topological2021yan}
    \bibinfo{author}{Yan, Z.}, \bibinfo{author}{Wang, Y.-C.}, \bibinfo{author}{Ma,
      N.}, \bibinfo{author}{Qi, Y.} \& \bibinfo{author}{Meng, Z.~Y.}
    \newblock \bibinfo{title}{Topological phase transition and single/multi anyon
      dynamics of $\mathbb{Z}_2$ spin liquid}.
    \newblock \emph{\bibinfo{journal}{npj Quant. Mater.}}
      \textbf{\bibinfo{volume}{6}}, \bibinfo{pages}{39} (\bibinfo{year}{2021}).
    
    \bibitem{yan2022qdm}
    \bibinfo{author}{Yan, Z.}, \bibinfo{author}{Samajdar, R.},
      \bibinfo{author}{Wang, Y.-C.}, \bibinfo{author}{Sachdev, S.} \&
      \bibinfo{author}{Meng, Z.~Y.}
    \newblock \bibinfo{title}{Triangular lattice quantum dimer model with variable
      dimer density. {P}reprint at \url{https://arxiv.org/abs/2202.11100}}
      (\bibinfo{year}{2022}).
    
    \bibitem{yan2020improved}
    \bibinfo{author}{Yan, Z.}
    \newblock \bibinfo{title}{Improved sweeping cluster algorithm for quantum dimer
      model. {P}reprint at \url{https://arxiv.org/abs/2011.08457}}
      (\bibinfo{year}{2020}).
    
    \bibitem{spinice_1}
    \bibinfo{author}{Castelnovo, C.}, \bibinfo{author}{Moessner, R.} \&
      \bibinfo{author}{Sondhi, S.}
    \newblock \bibinfo{title}{Magnetic monopoles in spin ice}.
    \newblock \emph{\bibinfo{journal}{Nature}} \textbf{\bibinfo{volume}{451}},
      \bibinfo{pages}{42--45} (\bibinfo{year}{2008}).
    
    \bibitem{gang_chen}
    \bibinfo{author}{Chen, G.}
    \newblock \bibinfo{title}{Dirac's ``magnetic monopoles'' in pyrochlore ice
      ${U}(1)$ spin liquids: Spectrum and classification}.
    \newblock \emph{\bibinfo{journal}{Phys. Rev. B}} \textbf{\bibinfo{volume}{96}},
      \bibinfo{pages}{195127} (\bibinfo{year}{2017}).
    
    \bibitem{RK2021}
    \bibinfo{author}{Zhou, Z.}, \bibinfo{author}{Yan, Z.}, \bibinfo{author}{Liu,
      C.}, \bibinfo{author}{Chen, Y.} \& \bibinfo{author}{Zhang, X.-F.}
    \newblock \bibinfo{title}{Emergent rokhsar-kivelson point in realistic quantum
      ising models. {P}reprint at \url{https://arxiv.org/abs/2106.05518}}
      (\bibinfo{year}{2021}).
    
    \bibitem{frank}
    \bibinfo{author}{Fulde, P.} \& \bibinfo{author}{Pollmann, F.}
    \newblock \bibinfo{title}{Strings in strongly correlated electron systems}.
    \newblock \emph{\bibinfo{journal}{Ann. Phys. (Berlin)}}
      \textbf{\bibinfo{volume}{17}}, \bibinfo{pages}{441--449}
      (\bibinfo{year}{2008}).
    
    \bibitem{yan2022targeting}
    \bibinfo{author}{Yan, Z.}, \bibinfo{author}{Zhou, Z.}, \bibinfo{author}{Wang,
      Y.-C.}, \bibinfo{author}{Meng, Z.~Y.} \& \bibinfo{author}{Zhang, X.-F.}
    \newblock \bibinfo{title}{Targeting topological optimization problem: Sweeping
      quantum annealing. {P}reprint at \url{https://arxiv.org/abs/2105.07134}}
      (\bibinfo{year}{2022}).
    
    \bibitem{tfim_str_1}
    \bibinfo{author}{Jiang, Y.} \& \bibinfo{author}{Emig, T.}
    \newblock \bibinfo{title}{String picture for a model of frustrated quantum
      magnets and dimers}.
    \newblock \emph{\bibinfo{journal}{Phys. Rev. Lett.}}
      \textbf{\bibinfo{volume}{94}}, \bibinfo{pages}{110604}
      (\bibinfo{year}{2005}).
    
    \bibitem{tfim_str_2}
    \bibinfo{author}{Jiang, Y.} \& \bibinfo{author}{Emig, T.}
    \newblock \bibinfo{title}{Ordering of geometrically frustrated classical and
      quantum triangular {I}sing magnets}.
    \newblock \emph{\bibinfo{journal}{Phys. Rev. B}} \textbf{\bibinfo{volume}{73}},
      \bibinfo{pages}{104452} (\bibinfo{year}{2006}).
    
    \bibitem{ising_3d}
    \bibinfo{author}{Iqbal, N.} \& \bibinfo{author}{McGreevy, J.}
    \newblock \bibinfo{title}{Toward a 3d {I}sing model with a weakly-coupled
      string theory dual}.
    \newblock \emph{\bibinfo{journal}{SciPost Phys.}} \textbf{\bibinfo{volume}{9}},
      \bibinfo{pages}{19} (\bibinfo{year}{2020}).
    
    \bibitem{zaanen_89}
    \bibinfo{author}{Zaanen, J.} \& \bibinfo{author}{Gunnarsson, O.}
    \newblock \bibinfo{title}{Charged magnetic domain lines and the magnetism of
      high-${T}_{c}$ oxides}.
    \newblock \emph{\bibinfo{journal}{Phys. Rev. B}} \textbf{\bibinfo{volume}{40}},
      \bibinfo{pages}{7391--7394} (\bibinfo{year}{1989}).
    
    \bibitem{zaanen_96}
    \bibinfo{author}{Eskes, H.}, \bibinfo{author}{Grimberg, R.},
      \bibinfo{author}{van Saarloos, W.} \& \bibinfo{author}{Zaanen, J.}
    \newblock \bibinfo{title}{Quantizing charged magnetic domain walls: Strings on
      a lattice}.
    \newblock \emph{\bibinfo{journal}{Phys. Rev. B}} \textbf{\bibinfo{volume}{54}},
      \bibinfo{pages}{R724--R727} (\bibinfo{year}{1996}).
    
    \bibitem{Machida}
    \bibinfo{author}{{Machida}, K.}
    \newblock \bibinfo{title}{{Magnetism in La$_{2}$CuO$_{4}$ based compounds}}.
    \newblock \emph{\bibinfo{journal}{Physica C}} \textbf{\bibinfo{volume}{158}},
      \bibinfo{pages}{192--196} (\bibinfo{year}{1989}).
    
    \bibitem{zaanen_00}
    \bibinfo{author}{Zaanen, J.}
    \newblock \bibinfo{title}{Order out of disorder in a gas of elastic quantum
      strings in $2+1$ dimensions}.
    \newblock \emph{\bibinfo{journal}{Phys. Rev. Lett.}}
      \textbf{\bibinfo{volume}{84}}, \bibinfo{pages}{753--756}
      (\bibinfo{year}{2000}).
    
    \bibitem{zaanen_01}
    \bibinfo{author}{Zaanen, J.}, \bibinfo{author}{Osman, O.~Y.},
      \bibinfo{author}{Kruis, H.~V.}, \bibinfo{author}{Nussinov, Z.} \&
      \bibinfo{author}{Tworzydlo, J.}
    \newblock \bibinfo{title}{The geometric order of stripes and {L}uttinger
      liquids}.
    \newblock \emph{\bibinfo{journal}{Phil. Mag. B}} \textbf{\bibinfo{volume}{81}},
      \bibinfo{pages}{1485--1531} (\bibinfo{year}{2001}).
    
    \bibitem{spinice_2}
    \bibinfo{author}{Gingras, M. J.~P.} \& \bibinfo{author}{McClarty, P.~A.}
    \newblock \bibinfo{title}{Quantum spin ice: a search for gapless quantum spin
      liquids in pyrochlore magnets}.
    \newblock \emph{\bibinfo{journal}{Rep. Prog. Phys.}}
      \textbf{\bibinfo{volume}{77}}, \bibinfo{pages}{056501}
      (\bibinfo{year}{2014}).
    
    \bibitem{spin_ice}
    \bibinfo{author}{Pomaranski, D.} \emph{et~al.}
    \newblock \bibinfo{title}{Absence of {P}auling's residual entropy in thermally
      equilibrated {D}y$_2${T}i$_2${O}$_7$}.
    \newblock \emph{\bibinfo{journal}{Nature Phys.}} \textbf{\bibinfo{volume}{9}},
      \bibinfo{pages}{353--356} (\bibinfo{year}{2013}).
    
    \bibitem{dirac_string}
    \bibinfo{author}{Morris, D. J.~P.} \emph{et~al.}
    \newblock \bibinfo{title}{Dirac strings and magnetic monopoles in the spin ice
      {D}y$_2${T}i$_2${O}$_7$}.
    \newblock \emph{\bibinfo{journal}{Science}} \textbf{\bibinfo{volume}{326}},
      \bibinfo{pages}{411--414} (\bibinfo{year}{2009}).
    
    \bibitem{SC_6}
    \bibinfo{author}{J.~Birgeneau, R.}, \bibinfo{author}{Stock, C.},
      \bibinfo{author}{M.~Tranquada, J.} \& \bibinfo{author}{Yamada, K.}
    \newblock \bibinfo{title}{Magnetic neutron scattering in hole-doped cuprate
      superconductors}.
    \newblock \emph{\bibinfo{journal}{J. Phys. Soc. Jpn.}}
      \textbf{\bibinfo{volume}{75}}, \bibinfo{pages}{111003}
      (\bibinfo{year}{2006}).
    
    \bibitem{incom_1}
    \bibinfo{author}{Zhang, X.-F.}, \bibinfo{author}{Hu, S.},
      \bibinfo{author}{Pelster, A.} \& \bibinfo{author}{Eggert, S.}
    \newblock \bibinfo{title}{Quantum domain walls induce incommensurate supersolid
      phase on the anisotropic triangular lattice}.
    \newblock \emph{\bibinfo{journal}{Phys. Rev. Lett.}}
      \textbf{\bibinfo{volume}{117}}, \bibinfo{pages}{193201}
      (\bibinfo{year}{2016}).
    
    \bibitem{str_cold_1}
    \bibinfo{author}{Mazurenko, A.} \emph{et~al.}
    \newblock \bibinfo{title}{A cold-atom {F}ermi--{H}ubbard antiferromagnet}.
    \newblock \emph{\bibinfo{journal}{Nature}} \textbf{\bibinfo{volume}{545}},
      \bibinfo{pages}{462--466} (\bibinfo{year}{2017}).
    
    \bibitem{str_cold_2}
    \bibinfo{author}{Chiu, C.~S.} \emph{et~al.}
    \newblock \bibinfo{title}{String patterns in the doped {H}ubbard model}.
    \newblock \emph{\bibinfo{journal}{Science}} \textbf{\bibinfo{volume}{365}},
      \bibinfo{pages}{251--256} (\bibinfo{year}{2019}).
    
    \bibitem{dun2018quantum}
    \bibinfo{author}{Dun, Z.} \emph{et~al.}
    \newblock \bibinfo{title}{Quantum versus classical spin fragmentation in
      dipolar kagome ice {H}o$_3${M}g$_2${S}b$_3${O}$_{14}$}.
    \newblock \emph{\bibinfo{journal}{Phys. Rev. X}} \textbf{\bibinfo{volume}{10}},
      \bibinfo{pages}{031069} (\bibinfo{year}{2020}).
    
    \bibitem{tmgo_neu}
    \bibinfo{author}{Shen, Y.} \emph{et~al.}
    \newblock \bibinfo{title}{Intertwined dipolar and multipolar order in the
      triangular-lattice magnet {T}m{M}g{G}a{O}$_4$}.
    \newblock \emph{\bibinfo{journal}{Nat. Commun.}} \textbf{\bibinfo{volume}{10}},
      \bibinfo{pages}{4530} (\bibinfo{year}{2019}).
    
    \bibitem{tmgo_uud}
    \bibinfo{author}{Li, Y.} \emph{et~al.}
    \newblock \bibinfo{title}{Partial up-up-down order with the continuously
      distributed order parameter in the triangular antiferromagnet
      {T}m{M}g{G}a{O}$_4$}.
    \newblock \emph{\bibinfo{journal}{Phys. Rev. X}} \textbf{\bibinfo{volume}{10}},
      \bibinfo{pages}{011007} (\bibinfo{year}{2020}).
    
    \bibitem{tmgo_kt}
    \bibinfo{author}{Li, H.} \emph{et~al.}
    \newblock \bibinfo{title}{{K}osterlitz-{T}houless melting of magnetic order in
      the triangular quantum {I}sing material {T}m{M}g{G}a{O}$_4$}.
    \newblock \emph{\bibinfo{journal}{Nat. Commun.}} \textbf{\bibinfo{volume}{11}},
      \bibinfo{pages}{1111} (\bibinfo{year}{2020}).
    
    \bibitem{tmgo_model}
    \bibinfo{author}{Liu, C.}, \bibinfo{author}{Huang, C.-J.} \&
      \bibinfo{author}{Chen, G.}
    \newblock \bibinfo{title}{Intrinsic quantum {I}sing model on a triangular
      lattice magnet $\mathrm{Tm}\mathrm{Mg}\mathrm{Ga}{\mathrm{o}}_{4}$}.
    \newblock \emph{\bibinfo{journal}{Phys. Rev. Research}}
      \textbf{\bibinfo{volume}{2}}, \bibinfo{pages}{043013} (\bibinfo{year}{2020}).
    
    \bibitem{mat_lgt_1}
    \bibinfo{author}{Chen, G.}
    \newblock \bibinfo{title}{Intrinsic transverse field in frustrated quantum
      {I}sing magnets: Physical origin and quantum effects}.
    \newblock \emph{\bibinfo{journal}{Phys. Rev. Research}}
      \textbf{\bibinfo{volume}{1}}, \bibinfo{pages}{033141} (\bibinfo{year}{2019}).
    
    \bibitem{flux}
    \bibinfo{author}{Schlittler, T.}, \bibinfo{author}{Barthel, T.},
      \bibinfo{author}{Misguich, G.}, \bibinfo{author}{Vidal, J.} \&
      \bibinfo{author}{Mosseri, R.}
    \newblock \bibinfo{title}{Phase diagram of an extended quantum dimer model on
      the hexagonal lattice}.
    \newblock \emph{\bibinfo{journal}{Phys. Rev. Lett.}}
      \textbf{\bibinfo{volume}{115}}, \bibinfo{pages}{217202}
      (\bibinfo{year}{2015}).
    
    \bibitem{orland_1993}
    \bibinfo{author}{Orland, P.}
    \newblock \bibinfo{title}{Exact solution of a quantum model of resonating
      valence bonds on the hexagonal lattice}.
    \newblock \emph{\bibinfo{journal}{Phys. Rev. B}} \textbf{\bibinfo{volume}{47}},
      \bibinfo{pages}{11280--11290} (\bibinfo{year}{1993}).
    
    \bibitem{orland_1991}
    \bibinfo{author}{Orland, P.}
    \newblock \bibinfo{title}{Fermionic strings and the exact solution of edge
      models in three dimensions}.
    \newblock \emph{\bibinfo{journal}{Int. J. Mod. Phys. B}}
      \textbf{\bibinfo{volume}{05}}, \bibinfo{pages}{2401--2438}
      (\bibinfo{year}{1991}).
    
    \bibitem{orland_1992}
    \bibinfo{author}{Orland, P.}
    \newblock \bibinfo{title}{Exact solution of a quantum gauge magnet in 2 + 1
      dimensions}.
    \newblock \emph{\bibinfo{journal}{Nucl. Phys. B}}
      \textbf{\bibinfo{volume}{372}}, \bibinfo{pages}{635--653}
      (\bibinfo{year}{1992}).
    
    \bibitem{tfim_qmc_3}
    \bibinfo{author}{Isakov, S.~V.} \& \bibinfo{author}{Moessner, R.}
    \newblock \bibinfo{title}{Interplay of quantum and thermal fluctuations in a
      frustrated magnet}.
    \newblock \emph{\bibinfo{journal}{Phys. Rev. B}} \textbf{\bibinfo{volume}{68}},
      \bibinfo{pages}{104409} (\bibinfo{year}{2003}).
    
    \bibitem{referee_2}
    \bibinfo{author}{Lin, S.-Z.}, \bibinfo{author}{Kamiya, Y.},
      \bibinfo{author}{Chern, G.-W.} \& \bibinfo{author}{Batista, C.~D.}
    \newblock \bibinfo{title}{Stiffness from disorder in triangular-lattice {I}sing
      thin films}.
    \newblock \emph{\bibinfo{journal}{Phys. Rev. Lett.}}
      \textbf{\bibinfo{volume}{112}}, \bibinfo{pages}{155702}
      (\bibinfo{year}{2014}).
    
    \bibitem{qmc_1}
    \bibinfo{author}{Sandvik, A.~W.}
    \newblock \bibinfo{title}{Finite-size scaling of the ground-state parameters of
      the two-dimensional {H}eisenberg model}.
    \newblock \emph{\bibinfo{journal}{Phys. Rev. B}} \textbf{\bibinfo{volume}{56}},
      \bibinfo{pages}{11678--11690} (\bibinfo{year}{1997}).
    
    \bibitem{luttinger}
    \bibinfo{author}{Haldane, F. D.~M.}
    \newblock \bibinfo{title}{`{L}uttinger liquid theory' of one-dimensional
      quantum fluids. {I}. properties of the {L}uttinger model and their extension
      to the general 1d interacting spinless {F}ermi gas}.
    \newblock \emph{\bibinfo{journal}{J. Phys. C}} \textbf{\bibinfo{volume}{14}},
      \bibinfo{pages}{2585--2609} (\bibinfo{year}{1981}).
    
    \bibitem{qmc_2}
    \bibinfo{author}{Sandvik, A.~W.}
    \newblock \bibinfo{title}{Stochastic series expansion method for quantum
      {I}sing models with arbitrary interactions}.
    \newblock \emph{\bibinfo{journal}{Phys. Rev. E}} \textbf{\bibinfo{volume}{68}},
      \bibinfo{pages}{056701} (\bibinfo{year}{2003}).
    
    \bibitem{qmc_3}
    \bibinfo{author}{Avella, A.} \& \bibinfo{author}{Mancini, F.}
    \newblock \emph{\bibinfo{title}{Strongly correlated systems: numerical
      methods}}, vol. \bibinfo{volume}{176} (\bibinfo{publisher}{Springer, Berlin},
      \bibinfo{year}{2013}).
    
    \bibitem{sac_1}
    \bibinfo{author}{Sandvik, A.~W.}
    \newblock \bibinfo{title}{Stochastic method for analytic continuation of
      quantum {M}onte {C}arlo data}.
    \newblock \emph{\bibinfo{journal}{Phys. Rev. B}} \textbf{\bibinfo{volume}{57}},
      \bibinfo{pages}{10287--10290} (\bibinfo{year}{1998}).
    
    \bibitem{sac_2}
    \bibinfo{author}{Shao, H.} \emph{et~al.}
    \newblock \bibinfo{title}{Nearly deconfined spinon excitations in the
      square-lattice spin-$1/2$ {H}eisenberg antiferromagnet}.
    \newblock \emph{\bibinfo{journal}{Phys. Rev. X}} \textbf{\bibinfo{volume}{7}},
      \bibinfo{pages}{041072} (\bibinfo{year}{2017}).
    
    \bibitem{sac_3}
    \bibinfo{author}{Sandvik, A.~W.}
    \newblock \bibinfo{title}{Constrained sampling method for analytic
      continuation}.
    \newblock \emph{\bibinfo{journal}{Phys. Rev. E}} \textbf{\bibinfo{volume}{94}},
      \bibinfo{pages}{063308} (\bibinfo{year}{2016}).
    
    \bibitem{tfim_map_2}
    \bibinfo{author}{Zhang, X.-F.} \& \bibinfo{author}{Eggert, S.}
    \newblock \bibinfo{title}{Chiral edge states and fractional charge separation
      in a system of interacting bosons on a kagome lattice}.
    \newblock \emph{\bibinfo{journal}{Phys. Rev. Lett.}}
      \textbf{\bibinfo{volume}{111}}, \bibinfo{pages}{147201}
      (\bibinfo{year}{2013}).
    
    \bibitem{tfim_map_3}
    \bibinfo{author}{Wan, Y.} \& \bibinfo{author}{Tchernyshyov, O.}
    \newblock \bibinfo{title}{Quantum strings in quantum spin ice}.
    \newblock \emph{\bibinfo{journal}{Phys. Rev. Lett.}}
      \textbf{\bibinfo{volume}{108}}, \bibinfo{pages}{247210}
      (\bibinfo{year}{2012}).
    
    \bibitem{tfim_map_4}
    \bibinfo{author}{Wan, Y.}, \bibinfo{author}{Carrasquilla, J.} \&
      \bibinfo{author}{Melko, R.~G.}
    \newblock \bibinfo{title}{Spinon walk in quantum spin ice}.
    \newblock \emph{\bibinfo{journal}{Phys. Rev. Lett.}}
      \textbf{\bibinfo{volume}{116}}, \bibinfo{pages}{167202}
      (\bibinfo{year}{2016}).
    
    \bibitem{sl_expr_3}
    \bibinfo{author}{Shen, Y.} \emph{et~al.}
    \newblock \bibinfo{title}{Evidence for a spinon {F}ermi surface in a
      triangular-lattice quantum-spin-liquid candidate}.
    \newblock \emph{\bibinfo{journal}{Nature}} \textbf{\bibinfo{volume}{540}},
      \bibinfo{pages}{559--562} (\bibinfo{year}{2016}).
    
    \bibitem{liao2021phase}
    \bibinfo{author}{Da~Liao, Y.} \emph{et~al.}
    \newblock \bibinfo{title}{Phase diagram of the quantum ising model on a
      triangular lattice under external field}.
    \newblock \emph{\bibinfo{journal}{Phys. Rev. B}}
      \textbf{\bibinfo{volume}{103}}, \bibinfo{pages}{104416}
      (\bibinfo{year}{2021}).
    
    \bibitem{af_elec}
    \bibinfo{author}{Shen, S.-P.} \emph{et~al.}
    \newblock \bibinfo{title}{Quantum electric-dipole liquid on a triangular
      lattice}.
    \newblock \emph{\bibinfo{journal}{Nat. Commun.}} \textbf{\bibinfo{volume}{7}},
      \bibinfo{pages}{10569} (\bibinfo{year}{2016}).
    
    \bibitem{trap_ion}
    \bibinfo{author}{Britton, J.~W.} \emph{et~al.}
    \newblock \bibinfo{title}{Engineered two-dimensional {I}sing interactions in a
      trapped-ion quantum simulator with hundreds of spins}.
    \newblock \emph{\bibinfo{journal}{Nature}} \textbf{\bibinfo{volume}{484}},
      \bibinfo{pages}{489--492} (\bibinfo{year}{2012}).
    
    \bibitem{QC}
    \bibinfo{author}{King, A.~D.} \emph{et~al.}
    \newblock \bibinfo{title}{Observation of topological phenomena in a
      programmable lattice of 1,800 qubits}.
    \newblock \emph{\bibinfo{journal}{Nature}} \textbf{\bibinfo{volume}{560}},
      \bibinfo{pages}{456--460} (\bibinfo{year}{2018}).
    
\end{thebibliography}
\end{document}